\documentclass[amsmath,amssymb,reprint,aps,prd,eqsecnum,floatfix]{revtex4-2}

\usepackage[dvipsnames]{xcolor}
\usepackage[colorlinks=true,citecolor=blue]{hyperref}
\usepackage{graphicx}
\usepackage{dcolumn}
\usepackage{bm}
\usepackage{mathrsfs}
\usepackage{subcaption}
\usepackage{caption}
\usepackage{soul}

\begin{document}

\title{Vacuum polarization on topological black holes with Robin boundary conditions}

\author{Thomas Morley}

\email{TMMorley1@sheffield.ac.uk}

\affiliation{Consortium for Fundamental Physics,
	School of Mathematics and Statistics,
	The University of Sheffield,
	Hicks Building,
	Hounsfield Road,
	Sheffield. S3 7RH United Kingdom}

\author{Peter Taylor}

\email{Peter.Taylor@dcu.ie}

\affiliation{Centre for Astrophysics and Relativity, School of Mathematical Sciences,\\ Dublin City University, Glasnevin, Dublin 9, Ireland}

\author{Elizabeth Winstanley}

\email{E.Winstanley@sheffield.ac.uk}

\affiliation{Consortium for Fundamental Physics,
School of Mathematics and Statistics,
The University of Sheffield,
Hicks Building,
Hounsfield Road,
Sheffield. S3 7RH United Kingdom}

\date{\today}

\begin{abstract}
	We compute the renormalized vacuum polarization for a massless, conformally coupled scalar field on asymptotically anti-de Sitter black hole backgrounds. 
	Mixed (Robin) boundary conditions are applied on the spacetime boundary. 
	We consider black holes with nonspherical event horizon topology as well as spherical event horizons.
	The quantum scalar field is in the Hartle-Hawking state, and we employ Euclidean methods to calculate the renormalized expectation values.
	Far from the black hole, we find that the vacuum polarization approaches a finite limit, which is the same for all boundary conditions except Dirichlet boundary conditions.
\end{abstract}

\maketitle

\section{Introduction}
\label{sec:intro}

The renormalized expectation value of the stress-energy tensor operator (RSET) $\langle {\hat {T}}_{\mu \nu }\rangle _{\rm {ren}}$ of a quantum field is a quantity of primary interest in quantum field theory in curved spacetime.
In the associated semi-classical approximation to quantum gravity, the spacetime background is classical and matter fields are quantized on a fixed background.
The RSET determines the back-reaction of the quantum field on the spacetime geometry  
via the semi-classical Einstein equations. If one naively replaces the classical stress-energy tensor with the corresponding expectation value of the quantum stress-energy tensor, the semi-classical equations would be
\begin{equation}
\label{eq:SCEinstein}
G_{\mu \nu }+ \Lambda g_{\mu \nu }= 8\pi \langle {\hat {T}}_{\mu \nu }\rangle,
\end{equation}
where $G_{\mu \nu }$ is the classical Einstein tensor, $\Lambda $ the cosmological constant, $g_{\mu \nu }$ the metric tensor and throughout this paper we employ units in which $c=G=\hbar =k_{B}=1$. However, the right-hand side of Eq.~(\ref{eq:SCEinstein}) is ill-defined since, for example, if the only field present were a quantum scalar field ${\hat {\Phi }}$, the stress-energy tensor operator ${\hat {T}}_{\mu \nu }$ involves terms which are quadratic in an operator-valued distribution evaluated at a single spacetime point. In other words, the stress-energy tensor must be renormalized and it is this RSET $\langle \hat{T}_{\mu\nu}\rangle_{\textrm{ren}}$ that ought to appear in Eq.~(\ref{eq:SCEinstein}). The cost of this mapping $\langle \hat{T}_{\mu\nu}\rangle\to \langle \hat{T}_{\mu\nu}\rangle_{\textrm{ren}}$ is the introduction of quadratic curvature terms in the semi-classical equations (see, for example, Ref.~\cite{WaldBook:1994}). While the renormalization problem had been conceptually solved by DeWitt and Christensen \cite{DeWitt:1975ys, Christensen:1976vb}, numerical implementation of the renormalization prescription in black hole spacetimes is a practical challenge. Moreover, the RSET involves second order derivatives acting on the quantum field being considered, which adds to the complications involved in practical computations. 
It is therefore instructive to also consider simpler expectation values.
The simplest nontrivial expectation value for a quantum scalar field theory is the vacuum polarization (VP) $\langle {\hat {\Phi }}^{2}\rangle $, which does not involve any derivatives of the field, and which will be the focus of this paper. We have dropped the subscript $\langle\,\, \rangle_{\textrm{ren}}$ for typographical convenience.

Quantum effects play a major role in black hole physics due to the emission of Hawking radiation \cite{Hawking:1974rv,Hawking:1974sw}.
Several decades after the discovery of Hawking radiation, the computation of expectation values of observables on black hole backgrounds remains an active area of research.
In pioneering work of Candelas and Howard, the VP \cite{Candelas:1980zt,Candelas:1984pg} and RSET \cite{Howard:1984qp,Howard:1985yg} were calculated on a Schwarzschild black hole.
They employed Euclidean methods, and considered a massless quantum scalar field in the Hartle-Hawking state \cite{Hartle:1976tp}. 
Following their work, expectation values for other quantum fields on Schwarzschild spacetime were also found \cite{Anderson:1989vg,Elster:1984hu,Fawcett:1983dk,Jensen:1988rh,Jensen:1992mv,Jensen:1995qv}.
Anderson, Hiscock and Samuel (AHS) \cite{Anderson:1993if,Anderson:1994hg} developed a general methodology for computing both the VP and RSET for a quantum scalar field with arbitrary mass and coupling to the spacetime curvature, on a static, spherically symmetric black hole background.
Their method was subsequently refined by Breen and Ottewill \cite{Breen:2010ux,Breen:2011af,Breen:2011aa} and has been applied to a variety of spherically symmetric black holes in four spacetime dimensions \cite{Breen:2015hwa,Flachi:2008sr,Quinta:2016eql,Winstanley:2007tf}.
There has also been more limited work on alternative approaches for nonspherically symmetric black hole space-times for which the AHS method is not applicable \cite{DeBenedictis:1998be,Ottewill:2010bq,Ottewill:2010hr,Ferreira:2014ina}.

In the past five years, new approaches to computations of the VP and RSET for a quantum scalar field have been developed \cite{Levi:2015eea,Taylor:2016edd,Freitas:2018mlu}.
The ``extended coordinates'' method of Taylor and Breen \cite{Taylor:2016edd,Taylor:2017sux,Breen:2018ukd}, like the AHS method, involves a Euclideanized spacetime and will be the method adopted here.
In contrast, the ``pragmatic mode-sum regularization'' scheme of Levi, Ori and collaborators works on the original Lorentzian black hole spacetime, and has been successfully applied to quantum scalar fields in a variety of quantum states on both static and stationary asymptotically flat black hole spacetimes \cite{Levi:2015eea,Levi:2016paz,Levi:2016quh,Levi:2016esr,Lanir:2018vgb,Lanir:2018rap,Zilberman:2019buh}.

With the notable exception of \cite{Flachi:2008sr,Ferreira:2014ina,Quinta:2016eql,Breen:2018ukd}, most of the works cited above concerned either asymptotically flat or asymptotically de Sitter black holes.
Black holes which are asymptotically anti-de Sitter (adS) are important within the context of the adS/CFT correspondence 
(see, for example, \cite{Aharony:1999ti} for a review).
Amongst asymptotically adS black holes, the three-dimensional BTZ black hole \cite{Banados:1992wn,Banados:1992gq,Carlip:1995qv} has received a great deal of attention in the literature.
The fact that the geometry of the BTZ spacetime is locally adS enables closed-form expressions to be found for renormalized expectation values \cite{Steif:1993zv,Lifschytz:1993eb,Shiraishi:1993nu,Shiraishi:1993ti,Binosi:1998yu},
in contrast to the four-dimensional situation, where numerical computations are required.
As a result, the back-reaction can be investigated explicitly in this case \cite{Martinez:1996uv,Casals:2016ioo,Casals:2016odj,Casals:2018ohr,Casals:2019jfo,Dias:2019ery,Emparan:2020rnp}.

A further richness in the theory of static, four-dimensional asymptotically adS black holes is that they do not necessarily have spherical event horizon topology (see, for example, \cite{Birmingham:1998nr,Brill:1997mf,Lemos:1994fn,Lemos:1994xp,Lemos:1995cm,Vanzo:1997gw,Cai:1996eg,Mann:1996gj,Smith:1997wx,Mann:1997zn}), in contrast to the situation for asymptotically flat black holes in four dimensions.
Moreover, while asymptotically flat Schwarzschild black holes are thermodynamically unstable, asymptotically adS black holes can be thermodynamically stable \cite{Hawking:1982dh,Brill:1997mf}, regardless of their event horizon topology.
For these reasons it is perhaps surprising that renormalized expectation values for quantum fields on four-dimensional asymptotically adS black hole backgrounds have 
not received more attention in the literature.

In \cite{Morley:2018lwn}, we studied the VP for a massless, conformally-coupled scalar field on topological black hole backgrounds.
Employing a Euclidean approach, we generalized the ``extended coordinates'' method of \cite{Taylor:2016edd,Taylor:2017sux,Breen:2018ukd} to black holes with flat or hyperbolic horizons and hence considered a field in the Hartle-Hawking state \cite{Hartle:1976tp}. 
The qualitative behaviour of the VP was similar for all event horizon topologies: the VP monotonically decreases from its value on the event horizon as the distance from the event horizon increases. 
Far from the black hole, the VP approaches a finite value equal to the vacuum expectation value of the VP in pure adS spacetime.

Quantum field theory on adS spacetime is complicated by the presence of a time-like boundary at null infinity, which means that adS is not globally hyperbolic. 
For this reason, to have a well-defined quantum field theory, it is necessary to apply boundary conditions to the field \cite{Avis:1977yn,Benini:2017dfw,Dappiaggi:2017wvj,Dappiaggi:2018pju,Dappiaggi:2018xvw,Ishibashi:2003jd,Ishibashi:2004wx,Wald:1980jn}.
In our previous work \cite{Morley:2018lwn}, we considered the simplest boundary conditions for a quantum scalar field, namely Dirichlet boundary conditions, for which the scalar field vanishes on the boundary.
However, Dirichlet boundary conditions are not the only possibility.
Very recently, we have studied quantum field theory on pure adS with general mixed (Robin) boundary conditions applied to the field \cite{Morley:2020ayr}.
The properties of the VP for both vacuum and thermal states depend on the particular boundary conditions applied: for some boundary conditions it is monotonically increasing from the origin to the boundary; for others monotonically decreasing. 
As the boundary is approached, the VP (for both vacuum and thermal states) tends to a finite limit, which again depends on the boundary conditions.
The value of the limit is the same for all boundary conditions other than Dirichlet, for which the limit takes a different value.
We therefore deduce that Dirichlet boundary conditions are rather nongeneric.

Inspired by our recent work on the effect of boundary conditions on quantum field theory in pure adS \cite{Morley:2020ayr}, in this paper we extend our previous study of the VP on topological black holes \cite{Morley:2018lwn} by considering general mixed (Robin) boundary conditions.
Using the methodology developed in \cite{Morley:2018lwn}, we compute the renormalized VP for a massless, conformally coupled scalar field on a variety of black holes with spherical, flat and hyperbolic horizons, paying particular attention to the effect of changing the boundary conditions satisfied by the scalar field.  

The outline of this paper is as follows.
In Sec.~\ref{sec:topBH} we review the classical properties of topological black holes, including their thermodynamics, before studying the classical behaviour of scalar field perturbations in Sec.~\ref{sec:classical}. 
The methodology for computing the renormalized VP is outlined in Sec.~\ref{sec:VP}, following \cite{Morley:2018lwn}. 
Our numerical results for the VP are presented in Sec.~\ref{sec:numerics}, while Sec.~\ref{sec:conc} contains our conclusions. 

\section{Topological black holes}
\label{sec:topBH}

Four-dimensional topological black holes are static solutions of the vacuum Einstein equations with negative cosmological constant. They are described by the metric \cite{Birmingham:1998nr,Brill:1997mf,Lemos:1994fn,Lemos:1994xp,Lemos:1995cm,Vanzo:1997gw,Cai:1996eg,Mann:1996gj,Smith:1997wx,Mann:1997zn}
\begin{equation}
ds^{2}=-f(r)dt^{2}+\frac{dr^{2}}{f(r)}+r^{2}d\Omega_{k}^{2}
\label{eq:metric}
\end{equation}
where $k$ can take the values $\{-1,0,1\}$, corresponding to negative, zero and positive horizon curvature respectively. 
The metric function $f(r)$ is given by
\begin{equation}
f(r)=k-\frac{2M}{r}+\frac{r^{2}}{L^{2}},
\label{eq:fdef}
\end{equation}
where $M$ is the black hole mass and $L$ is the adS curvature length-scale. The two-metric $d\Omega_{k}^{2}$ is defined by
\begin{equation}
d\Omega_{k}^{2}=d\theta^{2}+
\begin{cases}
\sin^{2}\theta  \, d\varphi^{2}, & k=1,\\ \theta^{2}\, d\varphi^{2}, & k=0, \\ \sinh^{2}\theta \, d\varphi^{2}, & k=-1.
\end{cases}
\end{equation}
For all $k$, the azimuthal coordinate $\varphi \in [0,2\pi )$.
The event horizon is located at $r=r_{h}$, which is the single real zero of the metric function $f(r)$ (\ref{eq:fdef}), so that $f(r_{h})=0$.

\begin{figure*}
	\centering
	\begin{subfigure}[b]{.45\textwidth}
		\centering\includegraphics[width=7cm]{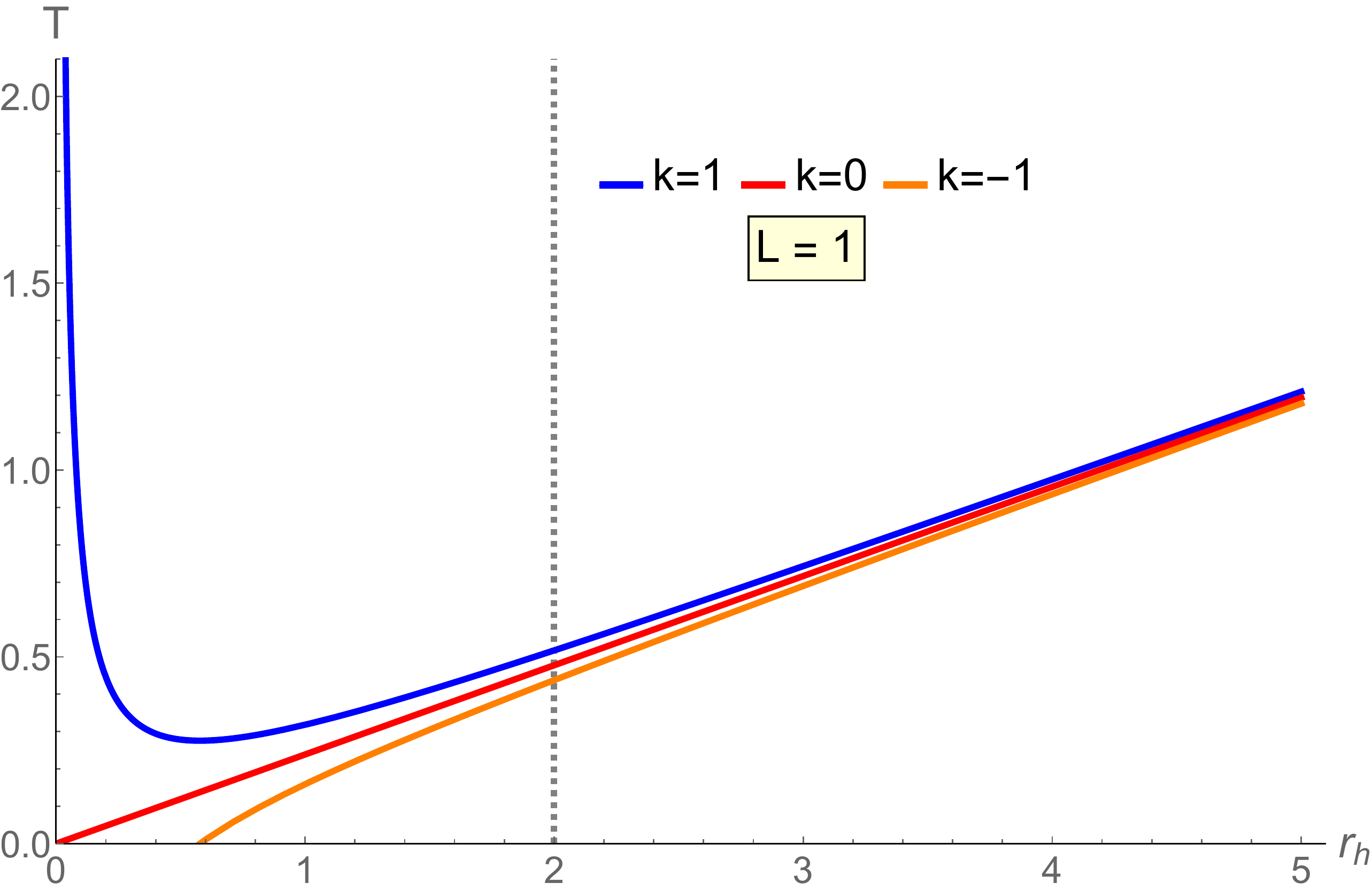}
		\subcaption{Topological black hole temperature $T$ (\ref{eq:temp}) as a function of event horizon radius $r_{h}$ for adS radius of curvature $L=1$. The intersections of the dotted line and the curves correspond to the three black holes with $r_{h}=2$ considered in Sec.~\ref{sec:numerics}.}
	\end{subfigure}
\begin{subfigure}[b]{.45\textwidth}
	\centering\includegraphics[width=7cm]{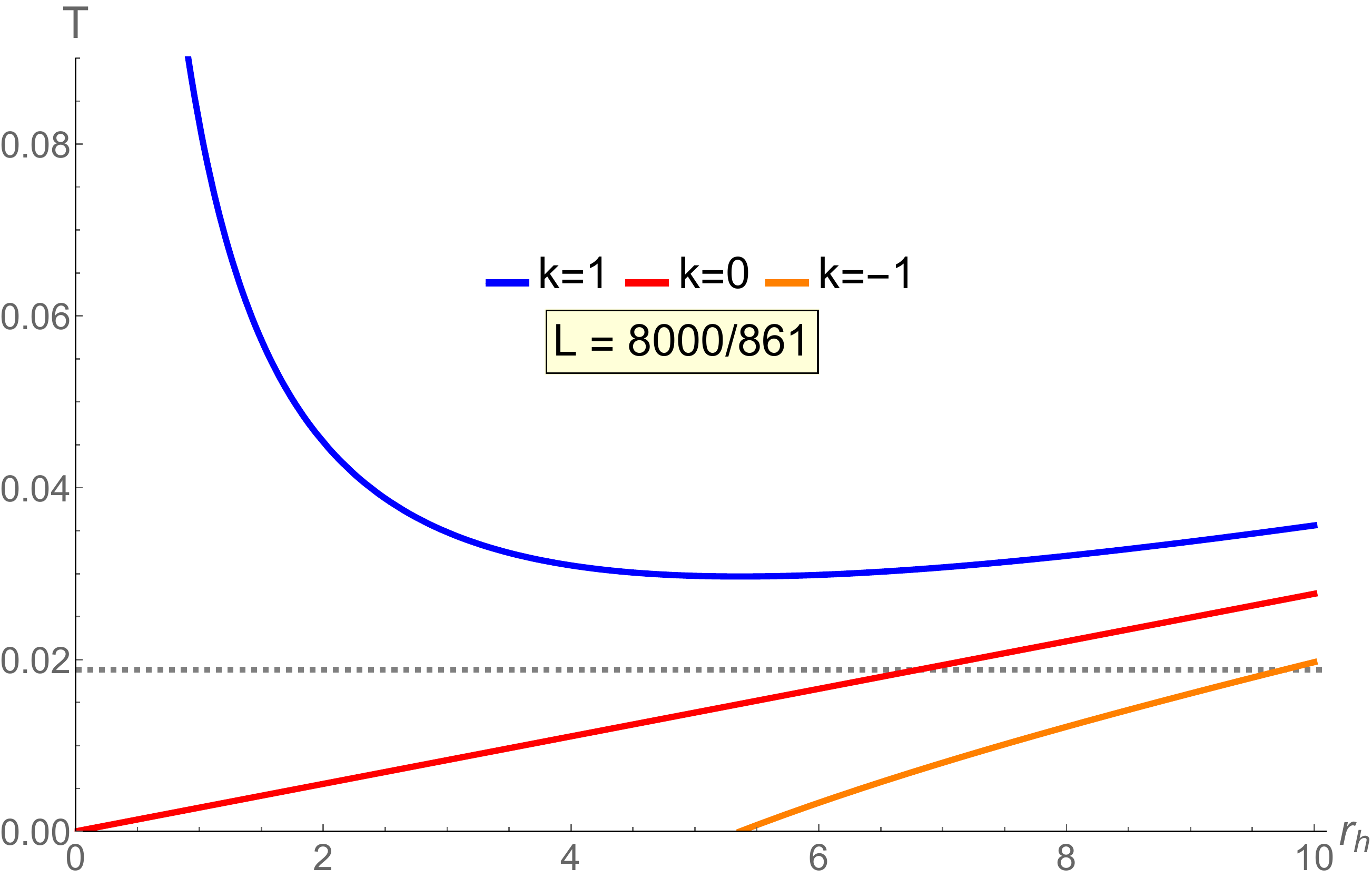}
	\subcaption{
	Topological black hole temperature $T$ (\ref{eq:temp}) as a function of event horizon radius $r_{h}$ for adS radius of curvature $L=8000/861$. The intersections of the dotted line and the curves correspond to the two black holes with $T=37843/640000\pi$ considered in Sec.~\ref{sec:numerics}.}
\end{subfigure}\vspace{10pt}
	\begin{subfigure}[b]{.45\textwidth}
		\centering\includegraphics[width=7cm]{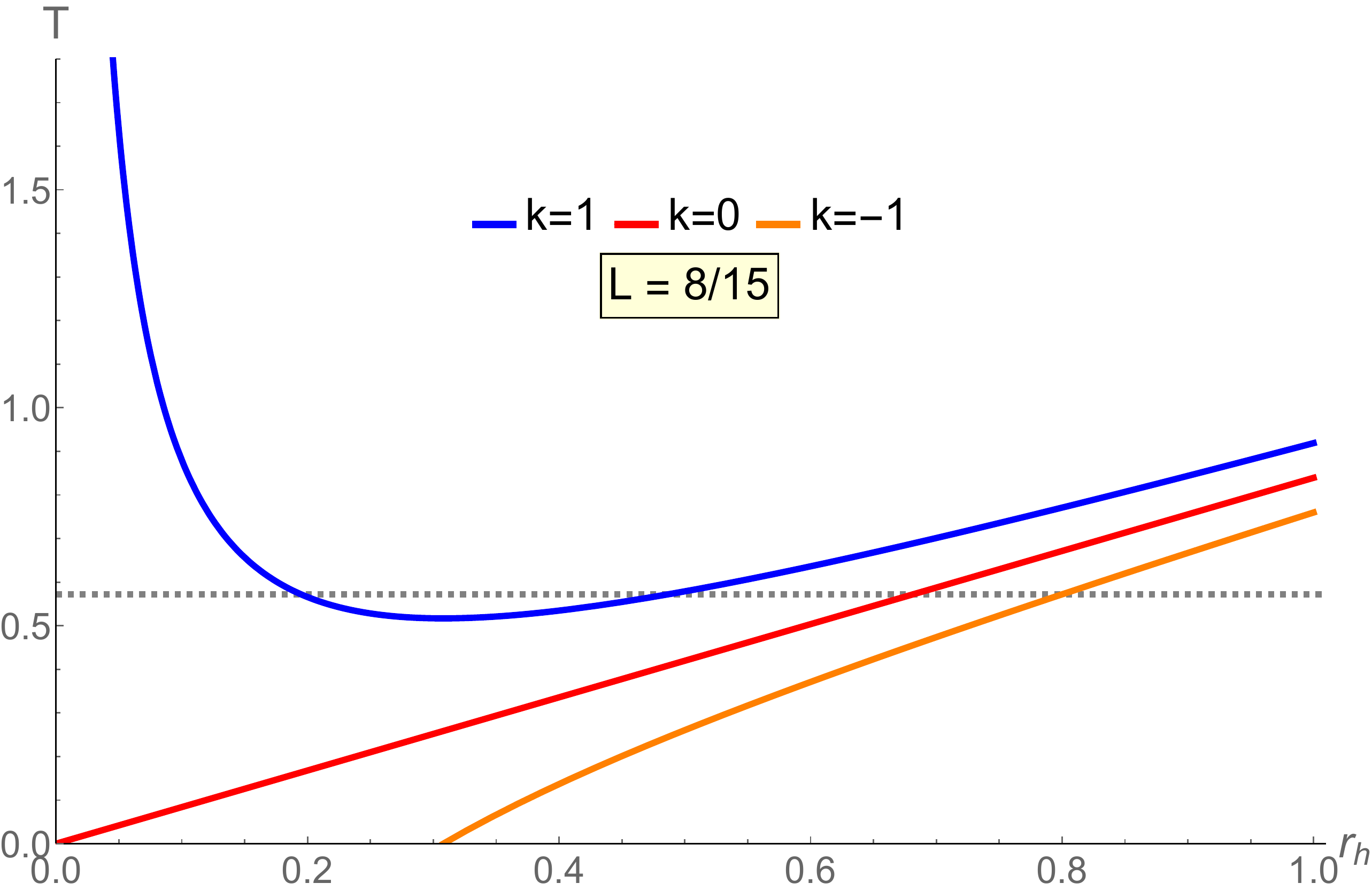}
		\subcaption{Topological black hole temperature $T$ (\ref{eq:temp}) as a function of event horizon radius $r_{h}$ for adS radius of curvature $L=8/15$. The intersections of the dotted line and the curves correspond to the four black holes with $T=115/64\pi$ considered in Sec.~\ref{sec:numerics}.}
	\end{subfigure}
	\begin{subfigure}[b]{.45\textwidth}
	\centering\includegraphics[width=7cm]{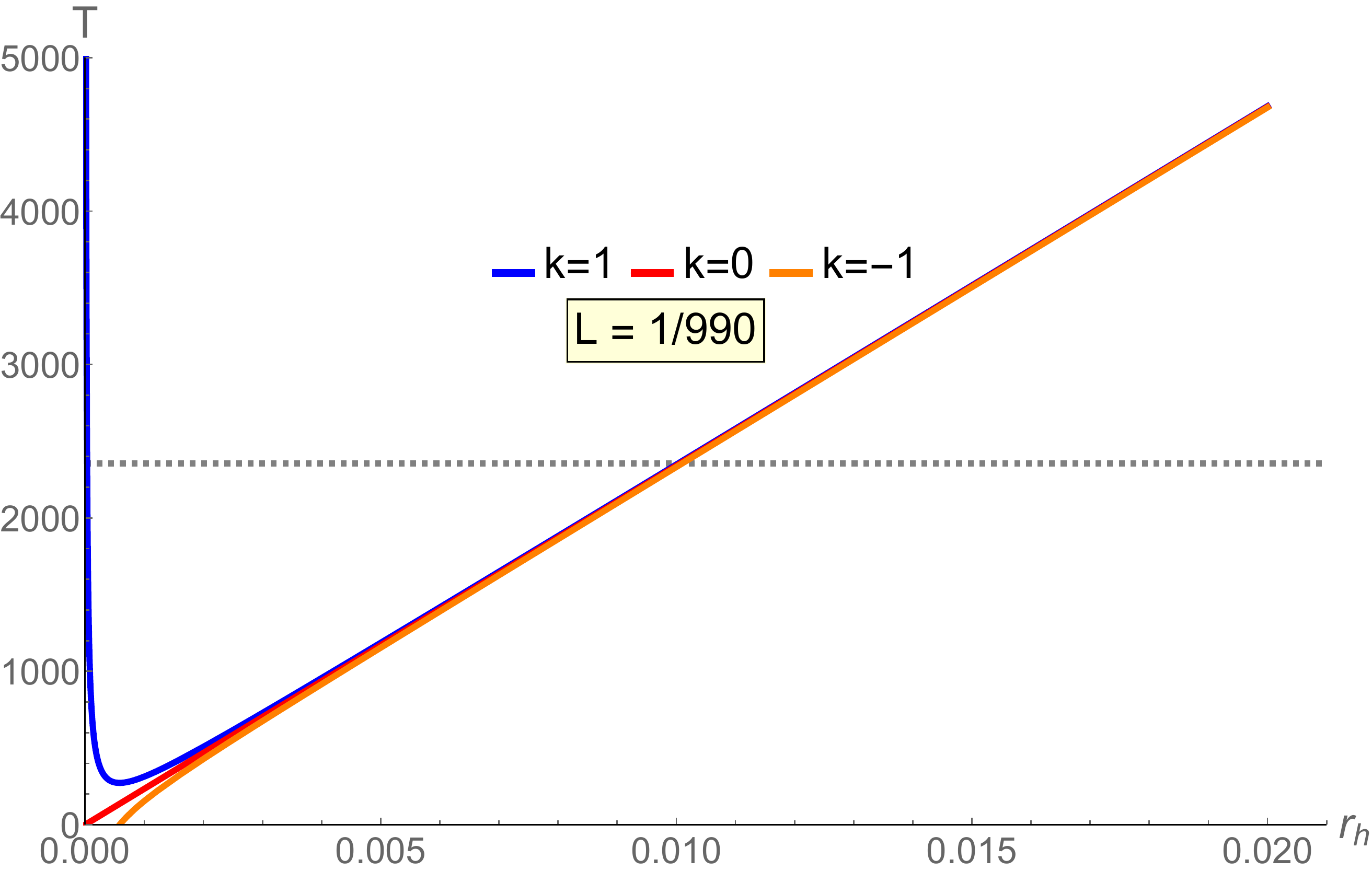}
	\subcaption{Topological black hole temperature $T$ (\ref{eq:temp}) as a function of event horizon radius $r_{h}$ for adS radius of curvature $L=1/990$. The intersections of the dotted line and the curves correspond to the four black holes with $T=29601/2\pi$ considered in Sec.~\ref{sec:numerics}.}
\end{subfigure}
\caption{Topological black hole temperature $T$ (\ref{eq:temp}) as a function of event horizon radius $r_{h}$ for fixed values of the adS radius of curvature $L$. The curves correspond to  $k=1$ spherical black holes (blue), $k=0$ planar black holes (red) and $k=-1$ hyperbolic black holes (orange).
In each plot the dotted line lies at the value of $r_{h}$ or $T$ corresponding to the particular black holes studied in Sec.~\ref{sec:numerics}. The curves are qualitatively the same in each of the four plots.}
\label{fig:temp}
\end{figure*}

When $k=1$ and the event horizon has constant positive curvature, we have the usual Schwarzschild-adS black hole. 
In this case $\theta \in [0,\pi ]$ is the usual spherical polar angle and the event horizon is a sphere. Black holes with $k=1$ exist for all positive values of the event horizon radius $r_{h}$.

When $k=0$ or $-1$, the event horizon is no longer compact (it is possible to form a compact horizon by making identifications \cite{Brill:1997mf}, but we do not consider this possibility here). In both cases, the range of the $\theta$ coordinate is $\theta\in[0,\infty)$.
For $k=0$, the event horizon has zero curvature and corresponds to a flat plane. In this case $\theta$ is the distance from a particular chosen origin in the plane.
Since the event horizon has vanishing curvature, there is only one length scale in the geometry, namely the adS radius of curvature $L$. 
As a result, the black hole metric (\ref{eq:metric}) has two scaling symmetries when $k=0$, which we can describe using an arbitrary constant $\rho $. 
First, there is the usual length rescaling
\begin{subequations}
	\label{eq:scaling}
\begin{equation}
t \rightarrow \rho t, \qquad r \rightarrow \rho r, \qquad M\rightarrow \rho M, \qquad L \rightarrow \rho L,
\label{eq:scaling1}
\end{equation}
which leaves $\theta $ invariant.
Second, we have
\begin{equation}
t\rightarrow \rho ^{-1}t, \qquad r\rightarrow \rho r, \qquad \theta \rightarrow \rho ^{-1}\theta , \qquad M \rightarrow \rho ^{3}M,
\label{eq:scaling2}
\end{equation}
\end{subequations}
with $L\rightarrow L$. 
All $k=0$ black holes are related, by the above scalings (\ref{eq:scaling}), to a chosen reference $k=0$ black hole spacetime.

For $k=-1$, the event horizon has constant negative curvature and hence is hyperbolic. Unlike the spherical and flat counterparts, there is a critical event horizon radius for the existence of hyperbolic black holes with
\begin{equation}
r_{h}> r_{h}^{\rm {crit}}:= \frac{L}{{\sqrt{3}}}.
\label{eq:rhmin}
\end{equation}

Topological black holes have a temperature $T$ given by 
\begin{equation}
T= \frac{\kappa }{2\pi } = \frac{f'(r_{h})}{4\pi } = \frac{kL^{2}+3r_{h}^{2}}{4\pi r_{h}L^{2}},
\label{eq:temp}
\end{equation}
where $\kappa =f'(r_{h})/2$ is the surface gravity of the black hole.
For $k=0$, $-1$, the temperature is a monotonically increasing function of horizon radius and black hole mass (see Fig.~\ref{fig:temp}), all black holes have positive specific heat and hence are thermodynamically stable \cite{Brill:1997mf}. 
Planar black holes with $k=0$ exist for all temperatures due to the scaling symmetries (\ref{eq:scaling}). 
Under the first scaling (\ref{eq:scaling1}), the temperature transforms as $T\rightarrow \rho ^{-1}T$, while under the second scaling (\ref{eq:scaling2}) we have
$T\rightarrow \rho T$.

When $k=1$, we see from Fig.~\ref{fig:temp} that there is also a minimum temperature $T_{\rm {min}}$ \cite{Hawking:1982dh}.
The minimum temperature for $k=1$ black holes occurs when $r_{h}=r_{h}^{\rm {crit}}$ (\ref{eq:rhmin}), and is given by 
\begin{equation}
T_{\rm {min}}= \frac{{\sqrt{3}}}{2\pi L}.
	\label{eq:Tmin}
\end{equation} 
Larger black holes have positive specific heat and are thermodynamically stable, while smaller black holes are thermodynamically unstable and have negative specific heat \cite{Hawking:1982dh}. 
For a fixed temperature $T>T_{\rm {min}}$, there are two black holes having the same temperature; one larger and one smaller.
We use the notation $k=1^{(+)}$ to denote larger, thermodynamically stable $k=1$ black holes and the notation $k=1^{(-)}$ to denote smaller, thermodynamically unstable $k=1$ black holes having the same temperature.

As a final note in this section, we will find it convenient for later use to introduce a dimensionless radial coordinate $\zeta$, defined by
\begin{equation}
\zeta = \frac{4\mu ^{2}+k}{M}r -1,
\label{eq:zetadef}
\end{equation}
where $\mu $ is a dimensionless parameter defined by 
\begin{equation}
\mu L \left( 4\mu ^{2}+k \right) =M.
\label{eq:mudef}
\end{equation}
The utility of this particular dimensionless radial coordinate is that the event horizon is located at $\zeta = 1$ for all $M$ and $L$, which renders comparison of results for different black hole parameters particularly straightforward. Only two of the parameters $L$, $M$, $\mu$ are linearly independent by virtue of (\ref{eq:mudef}). Note also that for $k=0$, the coordinate $\zeta$ is invariant under the scalings (\ref{eq:scaling}).

\section{Classical conformal scalar field on topological black holes}
\label{sec:classical}

We consider a massless, conformally coupled scalar field $\Phi $ satisfying the 
Klein-Gordon equation 
\begin{equation}
\left[ \nabla _{\mu }\nabla ^{\mu } +\frac{2}{L^{2}}\right] \Phi =0. 
\label{eq:KG}
\end{equation}
Mode solutions of this equation take the form
\begin{equation}
\Phi _{\omega\lambda m}(t,r, \theta, \varphi )=  e^{-i\omega t}{\mathcal {N}}_{\omega \lambda } X_{\omega\lambda}(r)Z_{\lambda m}(\theta,\varphi) ,
\label{eq:mode}
\end{equation}
where $\lambda $ is a separation constant (whose values will be given below),
$m$ is an integer, ${\mathcal {N}}_{\omega \lambda }$ is a normalization constant, 
the radial function $X_{\omega \lambda } (r)$ satisfies the equation
\begin{equation}
\left\{\frac{d}{dr}\left(r^{2}f(r)\frac{d}{dr}\right)+\frac{\omega^{2}r^{2}}{f(r)}-\nu_{\lambda}+\frac{2r^{2}}{L^{2}}\right\}X_{\omega\lambda}(r)=0, 
\label{eq:radial}
\end{equation}
with the constant $\nu _{\lambda }$ given by 
\begin{equation}
\nu_{\lambda}=\left[\lambda+\frac{1}{4}k(k+1)\right]^{2}-\frac{1}{4}k,
\label{eq:nudef}
\end{equation}
and the angular function $Z_{\lambda m}(\theta, \varphi )$ takes the form
\begin{equation}
Z_{\lambda m}(\theta,\varphi)=
\begin{cases}
Y_{\lambda m}(\theta,\varphi), & k=1,\\
J_{m}(\lambda\theta) e^{im\varphi}, & k=0,\\
P_{-\frac{1}{2}+i\lambda}^{|m|}(\cosh\theta) e^{im\varphi}, & k=-1.
\end{cases}
\end{equation}
When $k=1$, the separation constant $\lambda = 0,1,2, \ldots $ and the angular function $Z_{\lambda m}(\theta, \varphi )$ is the usual spherical harmonic $Y_{\lambda m}(\theta, \varphi )$.
For $k=0$, $-1$, the separation constant $\lambda $ is a continuous variable taking all values in the interval $[0,\infty )$.
When $k=0$, the angular function $Z_{\lambda m}(\theta, \varphi )$ involves a Bessel function $J_{m}(\lambda \theta )$, while for $k=-1$ we have a conical (Mehler) function $P_{-\frac{1}{2}+i\lambda}^{|m|}(\cosh\theta)$. 

In order to have a well-defined quantum field theory, boundary conditions must be imposed on the radial function $X_{\omega\lambda}(r)$ as $r\rightarrow \infty $. 
We impose Robin (mixed) boundary conditions, in which a linear combination of the radial function and its derivative normal to the spacetime boundary vanishes.
The spacetime boundary at $r\rightarrow \infty $ is not formally part of the spacetime, so,  in order to impose Robin boundary conditions, following \cite{Morley:2020ayr}, we make a conformal transformation to the Einstein static universe (ESU). 
The conformal transformation affects the metric and scalar field as follows:
\begin{equation}
g_{\mu \nu }\rightarrow \Omega ^{2}g_{\mu \nu }, \qquad \Phi \rightarrow \Omega ^{-1}\Phi  ,
\end{equation}
where $\Omega $ is the appropriate conformal factor. 
Let ${\overline {\tau }}$ be the dimensionless time coordinate and ${\overline {r}}$ be the dimensionless radial coordinate on the ESU, in terms of which the ESU metric takes the form
\begin{equation}
ds^{2}= L^{2} \left[ -d{\overline {\tau }}^{2} + d{\overline {r}}^{2} + \sin ^{2}{\overline {r}} \left( d\theta ^{2} + \sin ^{2}\theta \, d\varphi ^{2}\right) 
\right] ,
\label{eq:ESUmetric}
\end{equation}
where $\theta $ and $\varphi $ are the usual spherical polar coordinates.
Comparing (\ref{eq:metric}, \ref{eq:ESUmetric}) for $k=1$, when $r\rightarrow \infty $, we have ${\overline {r}}\rightarrow \pi /2$,
\begin{equation}
\Omega \sim \frac{L}{r}\sim -\frac{r_{*}}{L},
\end{equation}
and
\begin{equation}
\frac{d{\overline {r}}}{dr} \sim \frac{L}{r^{2}}\sim \frac{1}{L\,f(r)}, 
\end{equation}
where we have introduced the usual ``tortoise'' coordinate $r_{*}$ defined by 
\begin{equation}
\frac{dr_{*}}{dr} = \frac{1}{f(r)}
\label{eq:rstar}
\end{equation}
with the integration constant chosen in such a way that $r_{*} \rightarrow 0$ as $r\rightarrow \infty$.

Robin boundary conditions are imposed at ${\overline {r}}=\pi /2$, and take the form
\begin{equation}
\left[ \Omega ^{-1} X_{\omega \lambda }(r) \right] \cos \alpha + \frac{d}{d{\overline {r}}} \left[ \Omega ^{-1} X_{\omega \lambda }(r) \right] \sin \alpha = 0, 
\label{eq:boundary1}
\end{equation}
where the boundary conditions are parameterized by an angle $\alpha \in [0,\pi )$. 
Setting $\alpha =0$ corresponds to Dirichlet boundary conditions, while Neumann boundary conditions are given by $\alpha = \pi /2$.
In terms of the ``tortoise'' coordinate $r_{*}$, the boundary conditions (\ref{eq:boundary1}) become
\begin{equation}
 {\widetilde {X}}_{\omega \lambda }  \cos \alpha + L\frac{d{\widetilde {X}}_{\omega \lambda }}{dr_{*}} \sin \alpha = 0, 
\label{eq:boundary2}
\end{equation}
where we have defined 
\begin{equation}
{\widetilde {X}}_{\omega \lambda }(r) = rX_{\omega \lambda }(r).
\end{equation}

Before we can consider a quantum scalar field, we need to examine whether the classical scalar field is stable, that is, whether there exist mode solutions of the Klein-Gordon equation (\ref{eq:KG}) which grow exponentially with time.
If we impose either Dirichlet or Neumann boundary conditions, the classical scalar field has no unstable modes because a massless, conformally coupled scalar field satisfies the Breitenlohner-Freedman bound \cite{Breitenlohner:1982bm,Breitenlohner:1982jf}.
The situation for general Robin boundary conditions is more complex.

In pure adS, while the initial-value problem for the evolution of a classical scalar field satisfying Robin boundary conditions is well-defined for all values of the angle $\alpha $ \cite{Ishibashi:2004wx,Warnick:2012fi}, there is a range of values of $\alpha \in (\alpha _{\rm {crit}}, \pi )$, for which the dynamics is unstable. 
There are also unstable scalar field modes on four-dimensional, spherically symmetric, Schwarzschild-adS black holes for a certain range of values of $\alpha $ \cite{Holzegel:2012wt,Araneda:2016ecy}.
In this section we examine whether this is also the case for topological black holes, following the analysis in \cite{Holzegel:2012wt,Araneda:2016ecy} for the spherically symmetric case. 

First, using the method of \cite{Araneda:2016ecy}, we show that there exists an $\alpha _{\rm {crit}}\in (\pi /2,\pi )$ such that 
there are unstable modes when $\alpha \in (\alpha _{\rm {crit}},\pi )$. 
We begin by writing the radial equation (\ref{eq:radial}) in terms of the ``tortoise'' coordinate $r_{*}$ (\ref{eq:rstar}):
\begin{equation}
-\frac{d^{2}{\widetilde {X}}_{\omega \lambda }}{dr_{*}^{2}} + V_{\lambda }(r){\widetilde{X}}_{\omega \lambda } = \omega^{2}{\widetilde{X}}_{\omega \lambda },
\label{eq:schr}
\end{equation}
where the potential $V_{\lambda }(r)$ is given by
\begin{equation}
V_{\lambda }(r)=f(r)\left(\frac{\nu_{\lambda}}{r^{2}}+\frac{2M}{r^{3}}\right) .
\label{eq:potential}
\end{equation}
We note that $V_{\lambda }(r)>0$ for all $r\in [r_{h},\infty )$. 
We multiply both sides of (\ref{eq:schr}) by ${\widetilde{X}}_{\omega \lambda }$ and integrate over $r_{*}\in (-\infty ,0]$
to give
\begin{multline}
\omega^{2}\int_{-\infty}^{0}\left| {\widetilde{X}}_{\omega \lambda } \right| ^{2} \, dr_{*}= 
\left[ -{\widetilde {X}}_{\omega \lambda } \frac{d{\widetilde{X}}_{\omega \lambda }}{dr_{*}}\right]_{-\infty}^{0}
\\
+\int_{-\infty}^{0}\left[V_{\lambda }(r)\left| {\widetilde{X}}_{\omega \lambda } \right| ^{2}+\left| \frac{d{\widetilde{X}}_{\omega \lambda }}{dr_{*}} \right|^{2} \right] \, dr_{*},
\label{eq:schr3}
\end{multline}
where we have performed an integration by parts.
The first term on the right-hand-side of (\ref{eq:schr3}) vanishes when Neumann and Dirichlet conditions are imposed on the boundary. 
Noting that both remaining integrals must be positive, we find that $\omega^{2}>0$ and so all classical scalar field modes must be stable in these cases as expected.

When we impose Robin conditions on the boundary, 
the boundary term on the right-hand-side of (\ref{eq:schr3}) does not vanish. 
Provided $\cos\alpha\neq0$, using the boundary conditions (\ref{eq:boundary2}) we can write
\begin{equation}
\omega^{2}\int_{-\infty}^{0}\left| {\widetilde {X}}_{\omega \lambda } \right|^{2} \, dr_{*}
=L\tan\alpha \left|
\frac{d{\widetilde{X}}_{\omega \lambda }}{dr_{*}}(0)\right|^{2} +
{\mathcal {J}} ,
\label{eq:stability1}
\end{equation}
where
\begin{equation}
{\mathcal {J}}
=\int_{-\infty}^{0}\left[
V_{\lambda }(r)\left| {\widetilde {X}}_{\omega \lambda } \right|^{2}+\left|\frac{d{\widetilde {X}}_{\omega \lambda }}{dr_{*}}\right| ^{2}\right]dr_{*}>0.
\end{equation}
When $\alpha \in [0,\pi/2)$, it is the case that $\tan \alpha >0$ and  therefore $\omega ^{2}>0$, giving stable modes.
For $\alpha \in (\pi /2,\pi )$, the right-hand-side of (\ref{eq:stability1}) is not necessarily positive.
In this case Proposition 1 of \cite{Araneda:2016ecy} applies. 
Using a variational method, evaluating the right-hand-side of (\ref{eq:stability1}) for a test function ${\widetilde {X}}_{\omega \lambda }(r_{*}) = \exp \left( -\left[ r_{*}\tan \alpha \right] /L \right) $, we find that this is negative for sufficiently large $| \tan \alpha |$, and hence the Schr\"odinger operator on the left-hand-side of (\ref{eq:schr}) has a spectrum containing negative eigenvalues $\omega ^{2}$.
Such negative values of $\omega ^{2}$ correspond to unstable modes of the scalar field.

Having shown that there exists an $\alpha _{\rm {crit}}$ such that there are unstable modes for $\alpha \in (\alpha _{\rm {crit}},\pi )$, we now use the method of \cite{Holzegel:2012wt} to determine the values of $\alpha _{\rm {crit}}$ for topological black holes.
We thus seek solutions of the Schr\"odinger equation (\ref{eq:schr}) when $\omega^{2}$ crosses zero. 
For a fixed black hole spacetime, the potential $V_{\lambda }(r)$ (\ref{eq:potential}) satisfies
$V_{\lambda }(r)>V_{0}(r)$ for each fixed $r\in [r_{h},\infty )$.
Therefore, in order to find $\alpha _{\rm {crit}}$, it is sufficient to consider the modes for which $\lambda =0$.

\begin{figure}[t!]
	\centering
	\includegraphics[width=.95\linewidth]{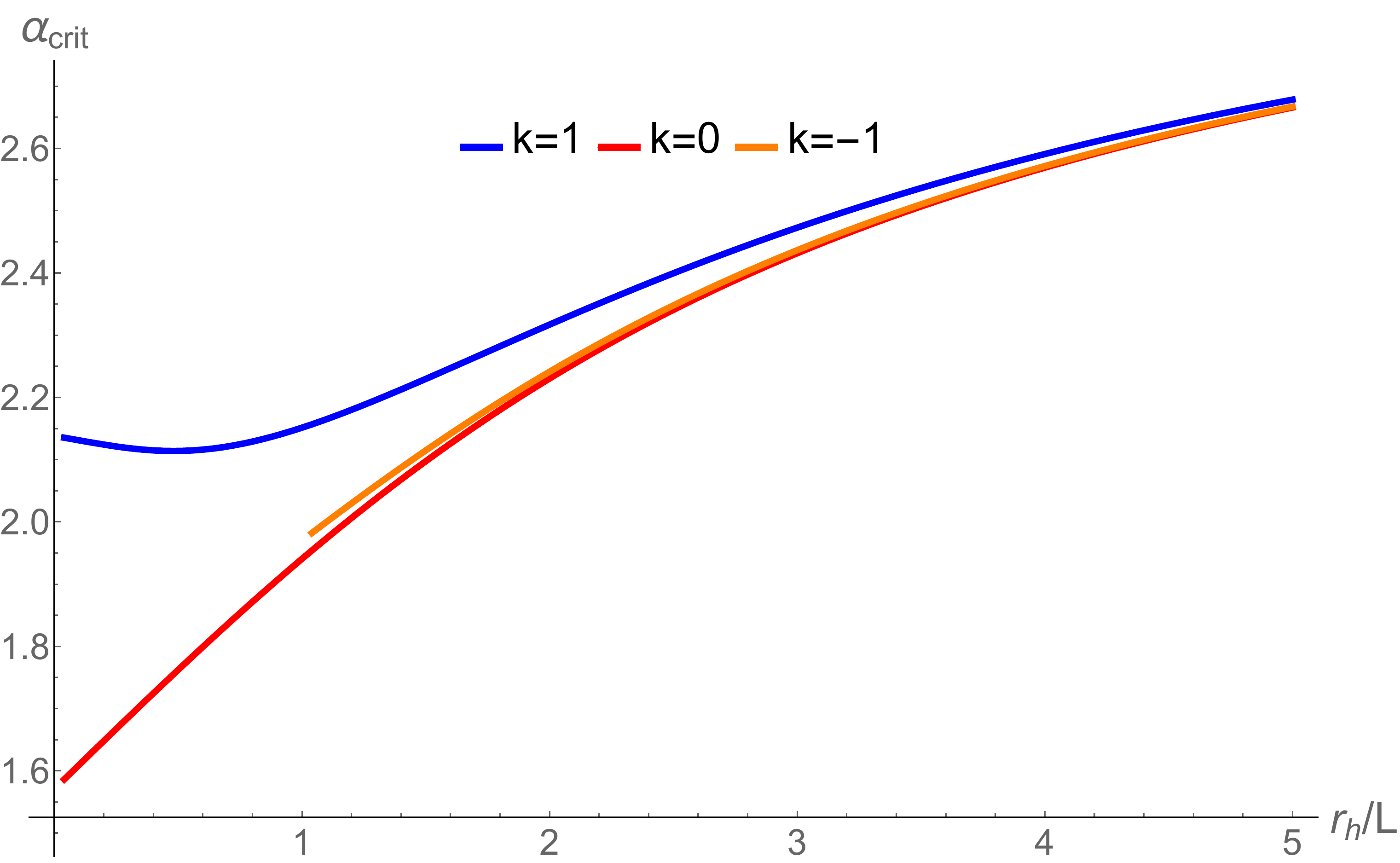}
	\caption{Critical value of $\alpha$ as a function of $r_{h}/L$ for spherically symmetric ($k=1$), planar ($k=0$) and hyperbolic ($k=-1$) black holes.
	Unstable classical scalar field modes exist for $\alpha >\alpha _{\rm {crit}}$.}
	\label{fig:alphacrit}
\end{figure}

When $\lambda = 0 = \omega ^{2}$, the radial equation (\ref{eq:radial}) takes the form
\begin{multline}
\frac{d}{dr}\left\{ \left[ r^{2}k-\left(kr_{h}+\frac{r_{h}^{3}}{L^{2}}\right)r+\frac{r^{4}}{L^{2}}\right] \frac{dX_{00}}{dr}\right\} 
\\+ \left[ -\nu_{0}+\frac{2r^{2}}{L^{2}} \right] X_{00}=0,
\end{multline}
where $\nu _{0}$ takes the values $\nu_{0}=0$ for $k=1,0$ and $\nu_{0}=\frac{1}{2}$ for $k=-1$, and  we have written the black hole mass $M$ in terms of the horizon radius $r_{h}$. 
We introduce dimensionless variables $R=r/L$, $R_{h}=r_{h}/L$, to give
\begin{multline}
\frac{d}{dR}\left\{ \left[ R^{2}k-(kR_{h}+R_{h}^{3})R+R^{4} \right]
\frac{dX_{00}}{dR} \right\}
\\
+\left[ -\nu_{0}+2R^{2}\right] X_{00}=0.
\label{eq:radialeq}
\end{multline}
For planar black holes with $k=0$, this equation takes the particularly simple form
\begin{equation}
\frac{d}{dR}\left\{ \left[ -R_{h}^{3}R+R^{4} \right]
\frac{dX_{00}}{dR} \right\}
+2R^{2}X_{00}=0,
\end{equation}
or equivalently, defining ${\overline {R}}=R/R_{h}$,
\begin{equation}
\frac{d}{d{\overline {R}}}\left\{ \left[ {\overline {R}}^{4}-{\overline {R}} \right]
\frac{dX_{00}}{d{\overline {R}}} \right\}
+2{\overline {R}}^{2}X_{00}=0.
\end{equation}
There is thus a single perturbation equation to be solved in the $k=0$ case.
This is to be expected, since all $k=0$ black hole spacetimes can be related via the scaling symmetries (\ref{eq:scaling}). 
The first scaling symmetry (\ref{eq:scaling1}) leaves $R$, $R_{h}$ (and hence ${\overline {R}}$) unchanged, while, under the second scaling symmetry (\ref{eq:scaling2}) we have $R\rightarrow \rho R$ and $R_{h}\rightarrow \rho R_{h}$, again leaving ${\overline {R}}$ unchanged.

To find $\alpha _{\rm {crit}}$, we solve (\ref{eq:radialeq}) numerically, integrating outwards from the horizon towards the spacetime boundary.
The value of $\alpha _{\rm {crit}}$ is then the value of $\alpha $ for which the derived solution $X_{00}(r)$ satisfies the boundary condition (\ref{eq:boundary2}), which, in terms of $R$ and $X_{00}$, takes the form
\begin{equation}
X_{00}(R) \cos \alpha _{\rm {crit}}+ R \frac{d}{dR} \left[ R X_{00}(R)\right] \sin \alpha _{\rm {crit}}= 0  
\label{eq:bcnumeric}
\end{equation} 
as $R\rightarrow \infty $.
To derive appropriate initial conditions for the numerical integration,
near the horizon, we define a new independent variable $x=R-R_{h}$ in terms of which  the radial equation (\ref{eq:radialeq}) takes the approximate form
\begin{equation}
\label{eq:bessels}
\left\{\frac{d^{2}}{dx^{2}}+\frac{1}{x}\frac{d}{dx}+\frac{2R_{h}^{2}-\nu_{0}}{(R_{h}k+3R_{h}^{3})x}\right\}X_{00}=0,
\end{equation}
where we have ignored higher order terms.
Solutions to (\ref{eq:bessels}) take the form of Bessel functions. 
Using these as initial conditions, we integrate out to a large value of $R$, and evaluate $\alpha _{\rm {crit}}$ from (\ref{eq:bcnumeric}). 

In Fig.~\ref{fig:alphacrit}, we show our numerical results for $\alpha _{\rm {crit}}$ for all three values of $k$.
The blue curve for spherically symmetric black holes ($k=1$) agrees with the results of Ref.~\cite{Holzegel:2012wt}.
As $r_{h}\rightarrow 0$, the value of $\alpha _{\rm {crit}}$ approaches the value in pure adS spacetime \cite{Morley:2020ayr}.
For $k=0$, the graph of $\cot \alpha _{\rm {crit}}$ as a function of $r_{h}/L$ is a straight line, as expected from the scaling symmetries (\ref{eq:scaling}) and the boundary condition (\ref{eq:bcnumeric}). 
For very small planar black holes with $k=0$, the value of $\alpha _{\rm {crit}}$ approaches $\pi /2$ as $r_{h}\rightarrow 0$.
The curve for hyperbolic black holes with $k=-1$ starts at the critical event horizon radius (\ref{eq:rhmin}) and lies very close to the curve for $k=0$ black holes.
The curves for $k=0$ and $k=-1$ are monotonically increasing as the event horizon radius increases, while that for $k=1$ is monotonically decreasing for small $r_{h}$, has a minimum and then monotonically increases for larger event horizon radius. 
As $r_{h}/L$ grows, the three curves merge.

\section{Vacuum polarization on topological black holes}
\label{sec:VP}

We now turn to the VP for a massless, conformally coupled quantum scalar field $\hat{\Phi }$ on topological black hole spacetimes. 
We follow the methodology of \cite{Morley:2018lwn}, which is reviewed briefly here.
Further details can be found in \cite{Morley:2018lwn}.

We begin by making a Wick rotation, setting $\tau= -it$, and work on the resulting Euclideanized spacetime. 
In order to avoid a conical singularity at the event horizon, the Euclidean ``time'' coordinate $\tau $ must be periodic with period $2\pi /\kappa $, where $\kappa $ is the surface gravity of the black hole.
Therefore we are considering a thermal state at the black hole temperature $T$ (\ref{eq:temp}), in other words the scalar field is in the Hartle-Hawking state \cite{Hartle:1976tp}.

The (unrenormalized) VP is defined as the coincidence limit of the Euclidean Green's function $G_{E}(x,x')$:
\begin{equation}
\langle \hat{\Phi }^{2}\rangle_{\rm {unren}}= \lim_{x'\rightarrow x} G_{E}(x,x') .
\label{eq:unren}
\end{equation}
The Euclidean Green's function can be found using standard separation of variables techniques and takes the form \cite{Morley:2018lwn}
\begin{equation}
    G_{E}(x,x')=\frac{\kappa}{4\pi^{2}}\sum_{n=-\infty}^{\infty}
    e^{in\kappa\Delta\tau}\int_{\lambda=0}^{\infty}d\lambda  \, \mathcal{P}_{\lambda}^{(k)}(\gamma)g_{n\lambda}^{\alpha }(r,r'),
    \label{eq:geuc}
\end{equation}
where 
\begin{equation}
    \mathcal{P}_{\lambda}^{(k)}(\gamma)=
    \begin{cases}
    (\lambda+\frac{1}{2})P_{\lambda}(\cos\gamma), & k=1,  \\
    \lambda J_{0}(\lambda\gamma), & k=0, \\
    \lambda  \tanh(\pi\lambda)P_{-\frac{1}{2}+i\lambda}(\cosh\gamma), & k=-1,
    \end{cases}
    \label{eq:calPdef}
\end{equation}
with $\gamma$ the geodesic distance on the two-surface with metric $d\Omega_{k}^{2}$, defined by
\begin{equation}
    \begin{array}{rcll}
        \cos\gamma &= &\cos\theta\cos\theta'+\sin\theta\sin\theta'\cos\Delta\phi, & k=1,  \\
        \gamma^{2}&= & \frac{1}{2}\left( \theta^{2}+\theta'^{2}-2\theta\theta'\cos\Delta\phi \right), & k=0,\\
        \cosh\gamma & = &\cosh\theta\cosh\theta'-\sinh\theta\sinh\theta'\cos\Delta\phi, & k=-1.
    \end{array}
\end{equation}
When $k=1$, the eigenvalue $\lambda$ is an integer and the integral in (\ref{eq:geuc}) should be replaced with a sum.
In (\ref{eq:calPdef}), $P_{\lambda }$ is a Legendre function, $J_{0}$ a Bessel function and $P_{-\frac{1}{2}+i\lambda }$ a conical function.

The radial Green's function $g_{n\lambda}^{\alpha }(r,r')$ satisfies the inhomogeneous equation
\begin{multline}
    \left\{\frac{d}{dr}\left(r^{2}f(r)\frac{d}{dr}\right)-\frac{n^{2}\kappa^{2}r^{2}}{f(r)}-\nu_{\lambda}+\frac{2r^{2}}{L^{2}}\right\}g_{n\lambda}^{\alpha }(r,r')
    \\ =-\delta(r-r').
    \label{eq:radialeuc}
\end{multline}
To solve (\ref{eq:radialeuc}), we write $g_{n\lambda}(r,r')$ as a normalised product of homogeneous solutions of the radial equation:
\begin{equation}
    g_{n\lambda}^{\alpha }(r,r')=\mathcal{C}_{n\lambda}^{\alpha }p_{n\lambda}(r_{<})q_{n\lambda }^{\alpha }(r_{>})
\end{equation}
where $r_{<}=\min\{r,r'\}$ and $r_{>}=\max\{r,r'\}$. The functions $p_{n\lambda}$ and $q_{n\lambda}^{\alpha }$ are both solutions of the homogeneous version of the radial equation (\ref{eq:radialeuc}).
The normalisation constant $\mathcal{C}_{n\lambda}^{\alpha }$ is constructed using the Wronskian $W\{p_{n\lambda}(r),q_{n\lambda}^{\alpha }(r)\}$ of these two functions:
\begin{equation}
    \mathcal{C}_{n\lambda}^{\alpha }=-\frac{1}{r^{2}f(r)W\{p_{n\lambda}(r),q_{n\lambda}^{\alpha }(r)\}}.
    \label{eq:normalization}
\end{equation}
The radial functions $p_{n\lambda }(r)$ are regular at the event horizon and do not depend on the angle $\alpha $ which parameterizes the Robin boundary conditions.

We impose Robin boundary conditions (\ref{eq:boundary2}) on the radial functions $q_{n\lambda }^{\alpha }(r)$ as $r\rightarrow \infty $.
We restrict attention to values of $\alpha $ in the interval $[0,\alpha _{\rm {crit}})$, for which the classical scalar field has no unstable modes. 
The radial functions satisfying Dirichlet ($q_{n\lambda }^{0}(r)$) and Neumann ($q_{n\lambda }^{\frac{\pi }{2}}(r)$) boundary conditions are linearly independent, and hence we may write, for any $\alpha \in [0,\alpha _{\rm {crit}})$
\begin{equation}
q_{n\lambda }^{\alpha }(r) = {\mathcal {A}}_{n\lambda }^{\alpha }q_{n\lambda }^{0}(r) + {\mathcal {B}}_{n\lambda }^{\alpha }q_{n\lambda }^{\frac{\pi }{2}}(r),
\end{equation}
where ${\mathcal {A}}_{n\lambda }^{\alpha }$ and ${\mathcal {B}}_{n\lambda }^{\alpha }$ are arbitrary constants. 
As $r\rightarrow \infty $, we normalize the Dirichlet and Neumann solutions by
\begin{equation}
q_{n\lambda }^{0}(r) = \frac{1}{r^{2}} + O(r^{-3}), \qquad q_{n\lambda }^{\frac{\pi }{2}}(r) = \frac{1}{r} + O(r^{-3}),
\end{equation}
with the overall constant set by the Wronskian condition. Substituting for $q_{n\lambda }^{\alpha }(r)$ in the boundary conditions (\ref{eq:boundary2}), the ratio of the constants ${\mathcal {A}}_{n\lambda }^{\alpha }$ and ${\mathcal {B}}_{n\lambda }^{\alpha }$ is fixed to be
\begin{equation}
\frac{{\mathcal {A}}_{n\lambda }^{\alpha }}{{\mathcal {B}}_{n\lambda }^{\alpha }}
= L \cot \alpha . 
\end{equation}
Without loss of generality, we take ${\mathcal {A}}_{n\lambda }^{\alpha }= L\cos \alpha $ and then
\begin{equation}
q_{n\lambda }^{\alpha } (r) = q_{n\lambda }^{0}(r) L\cos \alpha + q_{n\lambda }^{\frac{\pi }{2}}(r)  \sin \alpha .
\label{eq:qalpha}
\end{equation}
Using the linearity of the Wronskian, the normalization constants (\ref{eq:normalization}) can be written as
\begin{equation}
\mathcal{C}_{n\lambda}^{\alpha}=\frac{\mathcal{C}_{n\lambda}^{0}  \mathcal{C}_{n\lambda}^{\frac{\pi}{2}}}{ \mathcal{C}_{n\lambda}^{\frac{\pi}{2}} L \cos \alpha + \mathcal{C}_{n\lambda}^{0}\sin \alpha }.
\label{eq:normconstrob}
\end{equation}

The renormalized VP $\langle {\hat {\Phi }}^{2}\rangle $ is obtained from the unrenormalized expectation value (\ref{eq:unren}) by subtracting a Hadamard parametrix $G_{S}(x,x')$ for the geometric singular terms in the Euclidean Green's function and taking the coincidence limit, yielding:
\begin{equation}
\label{eq:VPDef}
\langle {\hat {\Phi }}^{2} \rangle = \lim _{x'\rightarrow x} \left[ G_{E}(x,x')- G_{S}(x,x') \right] .
\end{equation}
For a massless, conformally coupled scalar field, the Hadamard parametrix does not possess any logarithmic divergences and assumes a particularly simple form.
In this case the singular terms we need to consider are
\begin{equation}
G_{S}(x,x') = \frac{\Delta ^{\frac{1}{2}}}{8\pi ^{2}\sigma },
\label{eq:Gsing}
\end{equation}
where $\Delta $ is the Van Vleck-Morette determinant and $\sigma $ Synge's world function.
To facilitate numerical computation, a mode-sum representation of these singular terms is required so that the subtraction in Eq.~(\ref{eq:VPDef}) can be performed mode-by-mode.
This is constructed using the ``extended coordinates'' method of Refs.~\cite{Taylor:2016edd,Taylor:2017sux}, generalized to topological black holes in \cite{Morley:2018lwn}. 
First we take the partial coincidence limit by setting the radial separation of the space-time points $\Delta r=0$. 
We define new coordinates $w$ and $s$ by \cite{Morley:2018lwn}
\begin{equation}
\label{eq:ExpVarw}
w^{2}  = \frac{2}{\kappa^{2}}(1-\cos \kappa\Delta\tau),
\end{equation}
and
\begin{equation}
\label{eq:ExpVars}
s^{2} =  \begin{cases}
f(r)\,w^{2}+2 r^{2}(1-\cos\gamma), & k=1,\\
f(r)\,w^{2}+2 r^{2}\gamma^{2}, & k=0,\\
f(r)\,w^{2}+2 r^{2}(\cosh\gamma-1), & k=-1.
\end{cases}
\end{equation}
The singular terms (\ref{eq:Gsing}) are then expanded in terms of rational functions of $w$ and $s$ as follows:
\begin{equation}
\label{eq:HadamardDirectExp}
\frac{\Delta^{1/2}}{\sigma}= \sum_{i=0}^{2}\sum_{j=0}^{ i}\mathcal{D}_{ij}(r)\frac{w^{2i+2j}}{s^{2j+2}}
- \frac{f'(r)}{6r}
+\ldots ,
\end{equation}
where we have omitted terms in the expansion which vanish as the points are brought together. 
The coefficients ${\mathcal {D}}_{ij}(r)$ depend only on $k$, the surface gravity $\kappa $ and the metric function $f(r)$ and its derivatives. 
They are given explicitly in Table 1 in Ref.~\cite{Morley:2018lwn}.
\begin{widetext}
The mode-sum representation of the singular terms then results from writing the rational functions of $w$ and $s$ in (\ref{eq:HadamardDirectExp}) as  \cite{Morley:2018lwn}
\begin{equation}
\label{eq:RegParamAnsatz}
\frac{w^{2i+2j}}{s^{2j+2}}=\sum_{n=-\infty}^{\infty}e ^{i n \kappa\Delta\tau}\int_{\lambda=0}^{\infty}\mathcal{P}_{\lambda}^{(k)}(\gamma)\Psi_{n\lambda}^{(k)}(i,j|r) \, d \lambda ,
\end{equation}
where $\Psi^{(k)}_{\omega\lambda}(i,j|r)$ are regularization parameters given by
\cite{Morley:2018lwn}
\begin{equation}
	\Psi_{n\lambda}^{(k)}(i,j|r)=\frac{2^{i-j}i!(2i-1)!!(-1)^{n}}{\kappa^{2i+2j}r^{2j+2}j!}
	\sum_{p=n-i}^{n+i}\left(\frac{1}{\eta}\frac{\partial}{\partial\eta}\right)^{j}\frac{\chi_{p\lambda}^{(k)}(\eta)}{(i-n+p)!(i+n-p)!}
\end{equation}
with
\begin{eqnarray}
\chi_{p\lambda}^{(1)}(\eta) & = &
(-1)^{j}P_{\lambda}^{-|p|}(\eta)Q_{\lambda}^{|p|}(\eta),
\nonumber \\ 
\chi_{p\lambda}^{(0)}(\eta) & = &
\frac{1}{2}(-1)^{p+j}I_{p}(\lambda\eta)K_{p}(\lambda\eta) ,
\nonumber \\ 
\chi_{p\lambda}^{(-1)}(\eta) & = &
\frac{\pi}{2\cosh(\pi\lambda)}(-1)^{p}P_{-\frac{1}{2}+i\lambda}^{-p}(-\eta)P_{-\frac{1}{2}+i\lambda}^{p}(\eta) ,
\nonumber \\ & & 
\end{eqnarray}
and
\begin{equation}
\eta=\sqrt{\left|k+\frac{f(r)}{\kappa^{2}r^{2}}\right|}.
\end{equation}
Subtracting the resulting mode-sum representation of $G_{S}(x,x')$ from the Euclidean Green's function $G_{E}(x,x')$ (\ref{eq:geuc}) and bringing the separated points together gives the final expression for the VP:
\begin{equation}
\langle {\hat {\Phi }}^{2} \rangle  =  
\frac{1}{4\pi ^{2}}\int_{\lambda=0}^{\infty}d \lambda \sum_{n=-\infty}^{\infty}\,\mathcal{P}_{\lambda}^{(k)}(0)
\left[ \kappa g_{n\lambda}^{\alpha }(r )
-\frac{1}{2}\sum_{i=0}^{2}\sum_{j=0}^{i}\mathcal{D}_{ij}(r)\Psi_{n\lambda}^{(k)}(i,j|r)
\right]
-\frac{f'(r)}{48\pi^{2}r}, 
\label{eq:VPfinal}
\end{equation}
where we have defined
\begin{equation}
g_{n\lambda}^{\alpha }(r ) = \lim _{r'\rightarrow r} g_{n\lambda }^{\alpha }(r,r') = \frac{p_{n\lambda }(r)\mathcal{C}_{n\lambda}^{0}  \mathcal{C}_{n\lambda}^{\frac{\pi}{2}}}{ \mathcal{C}_{n\lambda}^{\frac{\pi}{2}} L\cos \alpha + \mathcal{C}_{n\lambda}^{0}\sin \alpha }\left[ q_{n\lambda }^{0}(r) L \cos \alpha + q_{n\lambda }^{\frac{\pi }{2}}(r)  \sin \alpha \right].
\end{equation}
\end{widetext}

\section{Numerical results}
\label{sec:numerics}

The renormalized VP (\ref{eq:VPfinal}) requires numerical computation.
Our methodology closely follows \cite{Morley:2018lwn}, so here we briefly summarize the key steps, and refer the reader to \cite{Morley:2018lwn} for more comprehensive details. 
All computations were carried out using Mathematica.

First the radial functions $p_{n\lambda }(r)$ are found by integrating the homogeneous version of the radial equation (\ref{eq:radialeuc}) from a point close to the event horizon out towards infinity.
Since we have the representation (\ref{eq:qalpha}) for the radial functions $q_{n\lambda }^{\alpha }(r)$, it is sufficient to find 
the functions $q_{n\lambda }^{0}(r)$ and $q_{n\lambda }^{\frac{\pi }{2}}(r)$ satisfying, respectively, Dirichlet and Neumann boundary conditions,
which significantly reduces computation time.
The functions $q_{n\lambda }^{0}(r)$ and $q_{n\lambda }^{\frac{\pi }{2}}(r)$ are  found by integrating  the homogeneous version of the radial equation (\ref{eq:radialeuc}) inwards from a large value of $r$. 
From these, the normalization constants ${\mathcal {C}}_{n\lambda }^{0,\frac{\pi }{2}}$ are computed using (\ref{eq:normalization}).
The analytic expressions for the coefficients ${\mathcal {D}}_{ij}(r)$ and the regularization parameters $\Psi _{n\lambda }(k)(i,j|r)$ (both of which are independent of the Robin angle $\alpha $) enable these to be straightforwardly computed.

We compute the VP for a range of values of $\alpha \in [0,\alpha _{\rm {crit}})$. 
Once $\alpha $ is fixed, we compute the sum over $n$ in (\ref{eq:VPfinal}) first. 
This converges extremely rapidly (see \cite{Morley:2018lwn} for more details of the convergence tests employed).
When $k=1$, we then have a sum over $\lambda =0,1,\ldots $, which also converges rapidly.
For $k=0$ and $-1$, the integral over $\lambda $ converges rapidly for large $\lambda $, and the major source of error in our final answers is the need to evaluate the integrand on a grid of values of $\lambda $ and use cubic spline interpolation between these grid points.
We use the same grid spacing as in \cite{Morley:2018lwn}, and estimate that the final relative error in the VP is no more than $\sim 10^{-3}$.

In \cite{Morley:2018lwn}, we presented plots of the VP for a selection of black holes, with Dirichlet boundary conditions applied to the scalar field. 
In this section, we present numerical results for general Robin boundary conditions and a selection of topological black hole spacetimes.
We begin by considering a set of topological black holes with event horizon radius $r_{h}=2$ and adS length-scale $L=1$, for comparison with results in \cite{Morley:2018lwn}. 
We have also chosen three additional sets of topological black hole solutions.
Within each set, we fix the adS length scale $L$ and the black hole surface gravity $\kappa $ (and hence temperature (\ref{eq:temp})). 
These additional black holes correspond to points in the $(r_{h},T)$-plane depicted in Fig.~\ref{fig:temp}.

The first additional set of solutions have a temperature which is below the minimum temperature $T_{\rm {min}}$ (\ref{eq:Tmin}) for spherical $k=1$ black holes, accordingly there are only planar $k=0$ and hyperbolic $k=-1$ black holes in this set.
In the remaining two sets of black holes, the temperature is above $T_{\rm {min}}$ and there are two $k=1$ black holes having the same temperature, as discussed in Sec.~\ref{sec:topBH}.
We consider a temperature just above $T_{\rm {min}}$, for which the two $k=1$ black holes are of a similar size, and also a very high temperature, when one of the $k=1$ black hole is much smaller than the other.

The numerical computations for hyperbolic $k=-1$ black holes are by far the most computationally intensive, due to the need to find a large number of modes in order to perform the integration over $\lambda $. 
For this reason, we have chosen the values of $L$ and $\kappa $ such that the $k=-1$ black holes correspond to those considered in \cite{Morley:2018lwn} with Dirichlet boundary conditions applied.
For planar black holes with $k=0$, we also need to compute an integral over $\lambda $ and hence a large number of modes, but we only need to find one set of modes, since all the $k=0$ black holes are related by the scalings (\ref{eq:scaling}). 
Finally, in our numerical results we have found it convenient to define a parameter $\beta$, related to the angle $\alpha $ in the boundary conditions (\ref{eq:boundary2}) by 
\begin{equation}
\cot \beta=L \cot \alpha .
\label{eq:betadef}
\end{equation}
Dirichlet and Neumann conditions are still recovered for $\beta=0$ and $\beta=\frac{\pi}{2}$ respectively.

\begin{table}
 \begin{tabular}{||c|| c| c | c | c||} 
 \hline &  & & & \\
  & $M$ & $r_{h}$ & $\kappa$ & $\beta_{{\rm {crit}}}$ \\ [0.5ex]
 \hline &  & & & \\
 $\mathbf{L=1}$  &  & & & \\
 $k=1$ & 5 & 2 & 3.25 & 2.3166 \\[0.1cm] 
 
 $k=0$ & 4 & 2 & 3 & 2.2301 \\[0.1cm]
 
 $k=-1$ & 3 & 2 & 2.75 & 2.2412 \\[1ex]
 \hline
  & & & &
 \\
 $\mathbf{L=\frac{8000}{861}}$ & & & & \\[0.15cm]
 $k=0$ & 1.8262 & 6.8064 & 0.1183 & 2.2300 \\[0.1cm]
 $k=-1$ & 0.5 & 9.7561 & 0.1183 & 2.9018 \\[1ex]
 \hline   & & & &
 \\
 $\mathbf{L=\frac{8}{15}}$ & & & & \\[0.15cm]
 $k=1^{(-)}$ & 0.1104 & 0.1948 & 3.5938 & 1.8833 \\[0.1cm]
 $k=1^{(+)}$ & 0.1104 & 0.4866 & 3.5938 & 1.8998 \\[0.1cm]
  $k=0$ & 0.5563 & 0.6815 & 3.5938 & 2.2301 \\[0.1cm]
  $k=-1$ & 0.5 & 0.8 & 3.5938 & 1.8829 \\[1ex]
  \hline   & & & &
  \\
  $\mathbf{L=\frac{1}{990}}$ & & & &\\[0.15cm]
  $k=1^{(-)}$ & 0.00002 & 0.00003 & 14800.5 & 1.5714 \\[0.1cm]
  $k=1^{(+)}$ & 0.00002 & 0.0101 & 14800.5 & 1.5747 \\[0.1cm]
  $k=0$ & 0.5000 & 0.0101 & 14800.5 & 2.2302 \\[0.1cm]
  $k=-1$ & 0.5 & 0.0101 & 14800.5 & 1.5747 \\[1ex] 
 \hline
\end{tabular}
\caption{Black hole parameters $L$, $\kappa $, $M$ and $r_{h}$ for which we present numerical results for the VP. 
We also give the critical value of the parameter $\beta$, related to $\alpha _{\rm {crit}}$ by (\ref{eq:betadef}). For values of $\beta $ greater than the critical value, there exist unstable classical scalar field modes.}
\label{tab}
\end{table}

Tab.~\ref{tab} details the various values of $L$, $M$ and $r_{h}$ for the black holes for which we have calculated the VP. 
Table~\ref{tab} also gives $\beta _{\rm {crit}}$, related to the critical angle 
$\alpha_{\rm {crit}}$ via (\ref{eq:betadef}) to four decimal places.
All the VP plots in the rest of this section will use the dimensionless radial coordinate $\zeta $ (\ref{eq:zetadef}), so that the event horizon is located at $\zeta =1$ for all black holes.

\begin{figure*}
	\centering
	\includegraphics[width=0.45\textwidth]{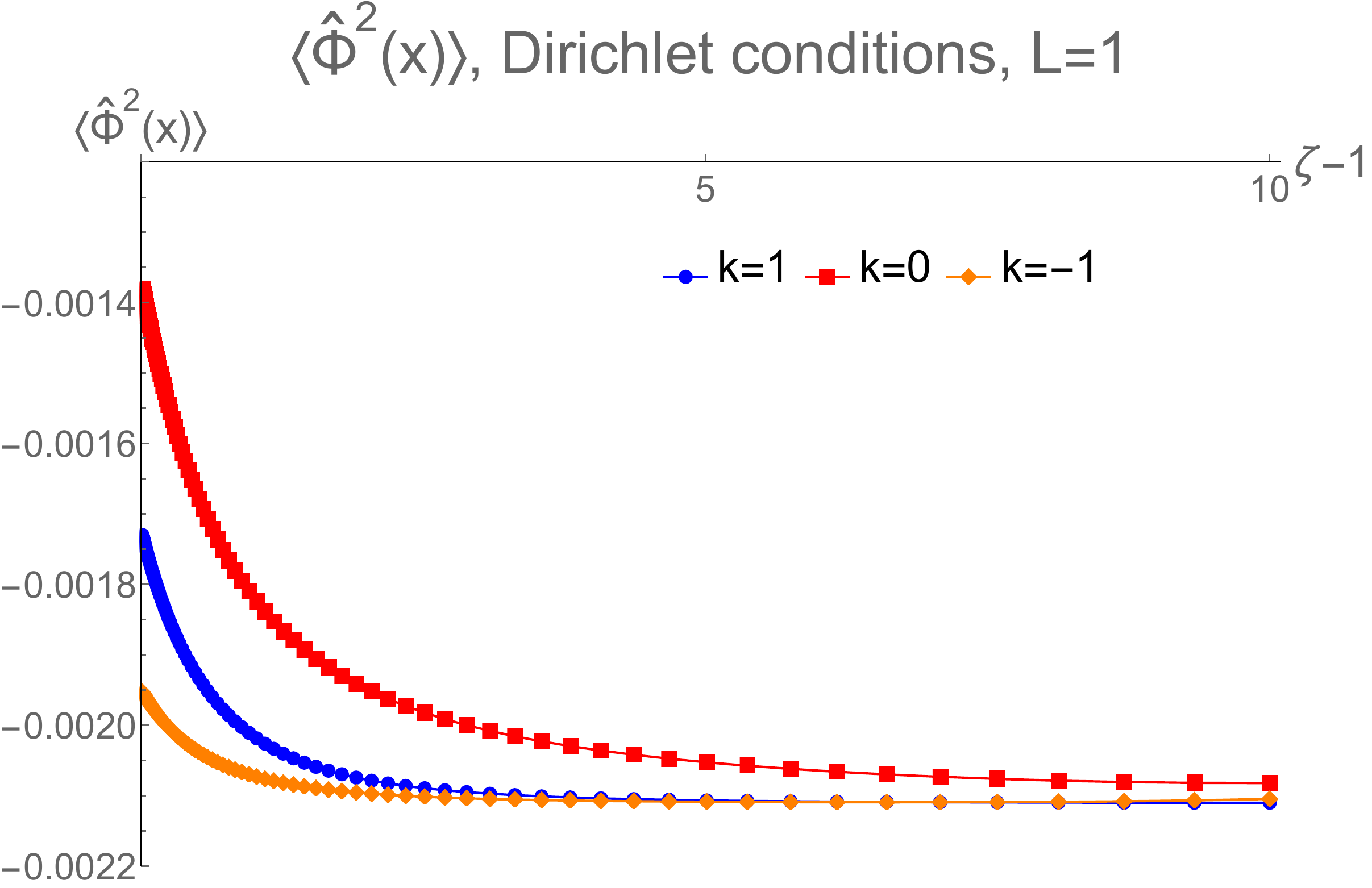}
	\includegraphics[width=0.45\textwidth]{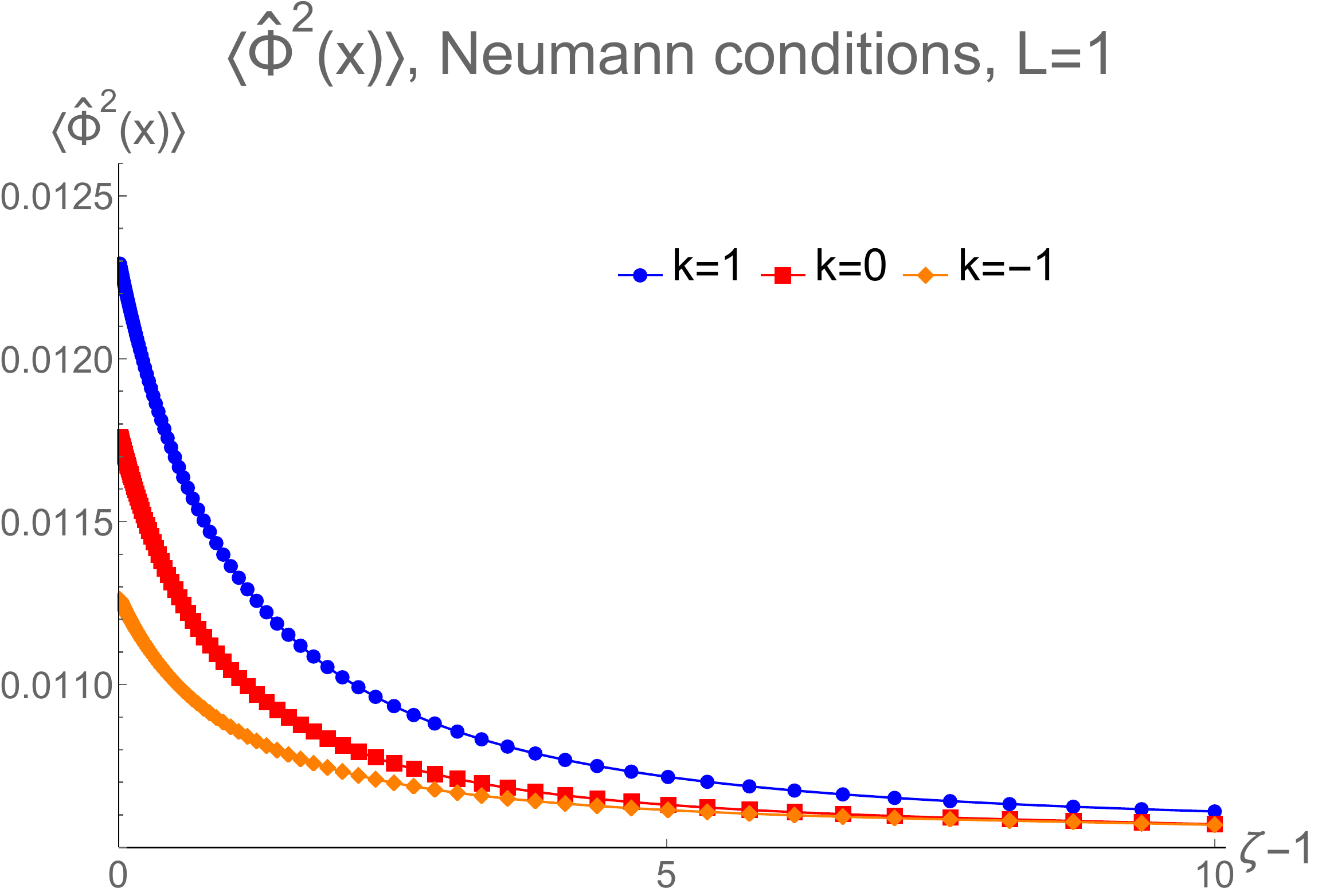}
	\caption{VP for topological black holes with adS radius of curvature $L=1$ and event horizon radius $r_{h}=2$, with Dirichlet (left) and Neumann (right) boundary conditions applied. For Dirichlet boundary conditions, the results are taken from \cite{Morley:2018lwn}.}
	\label{fig:L1rh2DirNeu}
\end{figure*}

We first consider topological black holes with $L=1$ and $r_{h}=2$. In Fig.~\ref{fig:L1rh2DirNeu}
we show the VP as a function of $\zeta $ with Dirichlet and Neumann boundary conditions applied.
The results for Dirichlet boundary conditions have previously appeared in \cite{Morley:2018lwn} and are repeated here for comparison.
For both Dirichlet and Neumann boundary conditions, and all values of $k$, the VP is monotonically decreasing from its value on the event horizon to its value on the spacetime boundary.
As $r\rightarrow \infty $, the VP approaches the vacuum value in pure adS for either Dirichlet boundary conditions,
\begin{subequations}
	\label{eq:adS}
\begin{equation}
\langle {\hat {\Phi }}^{2}\rangle _{\rm {adS,D}} = -\frac{1}{48\pi ^{2}L^{2}}, 
\label{eq:adSD}
\end{equation}
or Neumann boundary conditions
\begin{equation}
\langle {\hat {\Phi }}^{2}\rangle _{\rm {adS,N}} = \frac{5}{48\pi ^{2}L^{2}}, 
\label{eq:adSN}
\end{equation}
\end{subequations}
as applicable.
When Dirichlet boundary conditions are applied, the VP is negative everywhere on and outside the event horizon; in contrast, for Neumann boundary conditions the VP is positive everywhere for these particular black holes. 
The order of the curves in the two plots in Fig.~\ref{fig:L1rh2DirNeu} depends on the boundary conditions applied. For Dirichlet boundary conditions, the VP for $k=0$ black holes is greater than that for $k=1$ black holes, which in turn is greater than that for $k=-1$ black holes.
For Neumann boundary conditions, the $k=-1$ black holes have the smallest VP, followed by the $k=0$ black holes and then the $k=1$ black holes.
This suggests that the relative ordering between the $k=0$ and $k=1$ black holes may change for some intermediate value of $\beta $.  

\begin{figure*}
\centering
	\begin{subfigure}[b]{.45\textwidth}
		\centering\includegraphics[width=7cm]{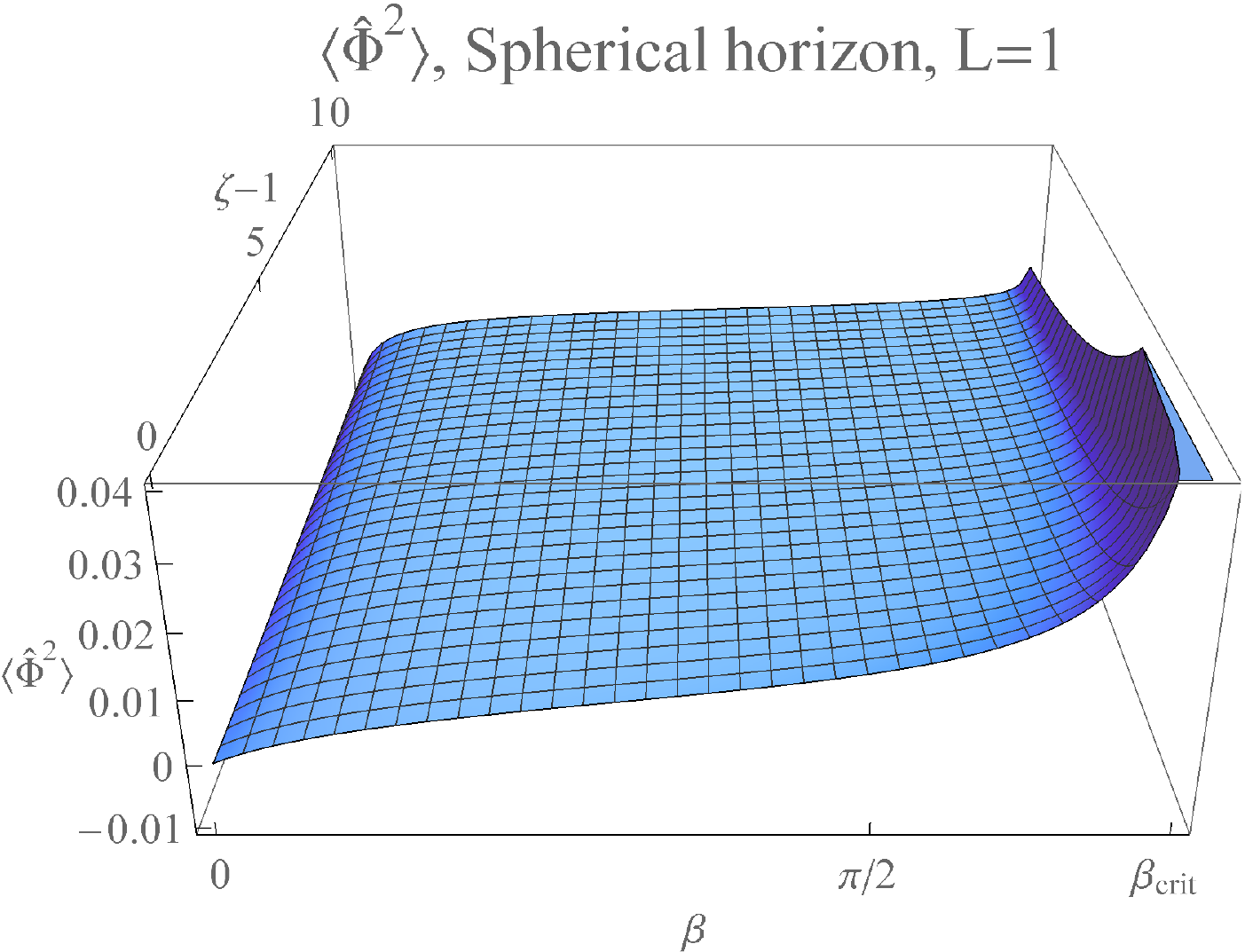}
		\subcaption{$k=1$, $M=5$, $\beta _{\rm {crit}}=2.3166$}
	\end{subfigure}
\begin{subfigure}[b]{.45\textwidth}
	\centering\includegraphics[width=7cm]{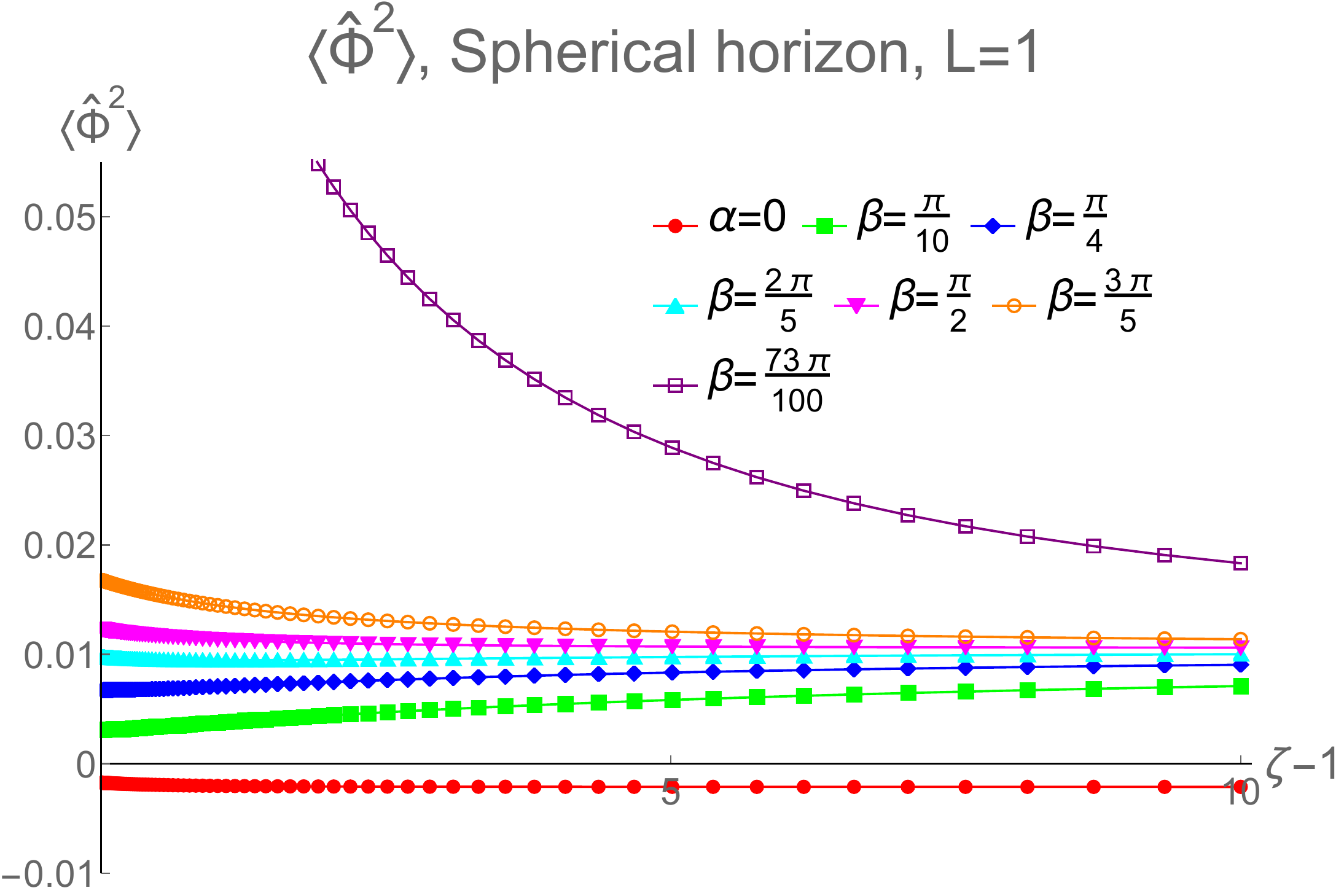}
	\subcaption{$k=1$, $M=5$, $\beta_{{\rm {crit}}}\simeq\frac{74\pi}{100}$}
\end{subfigure}\vspace{10pt}
	\begin{subfigure}[b]{.45\textwidth}
		\centering\includegraphics[width=7cm]{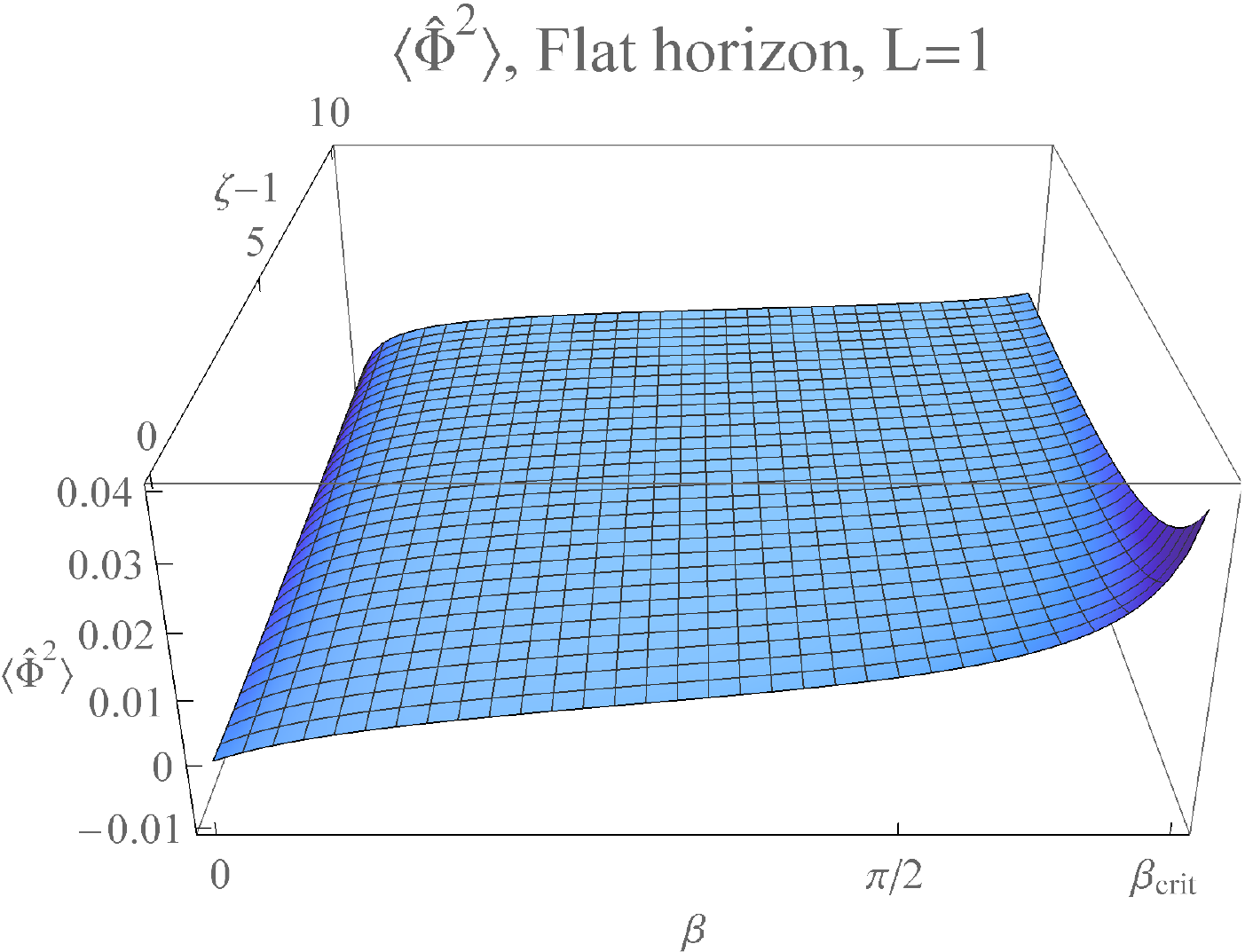}
		\subcaption{$k=0$, $M=4$, $\beta_{{\rm {crit}}}=2.2301$}
	\end{subfigure}
\begin{subfigure}[b]{.45\textwidth}
	\centering\includegraphics[width=7cm]{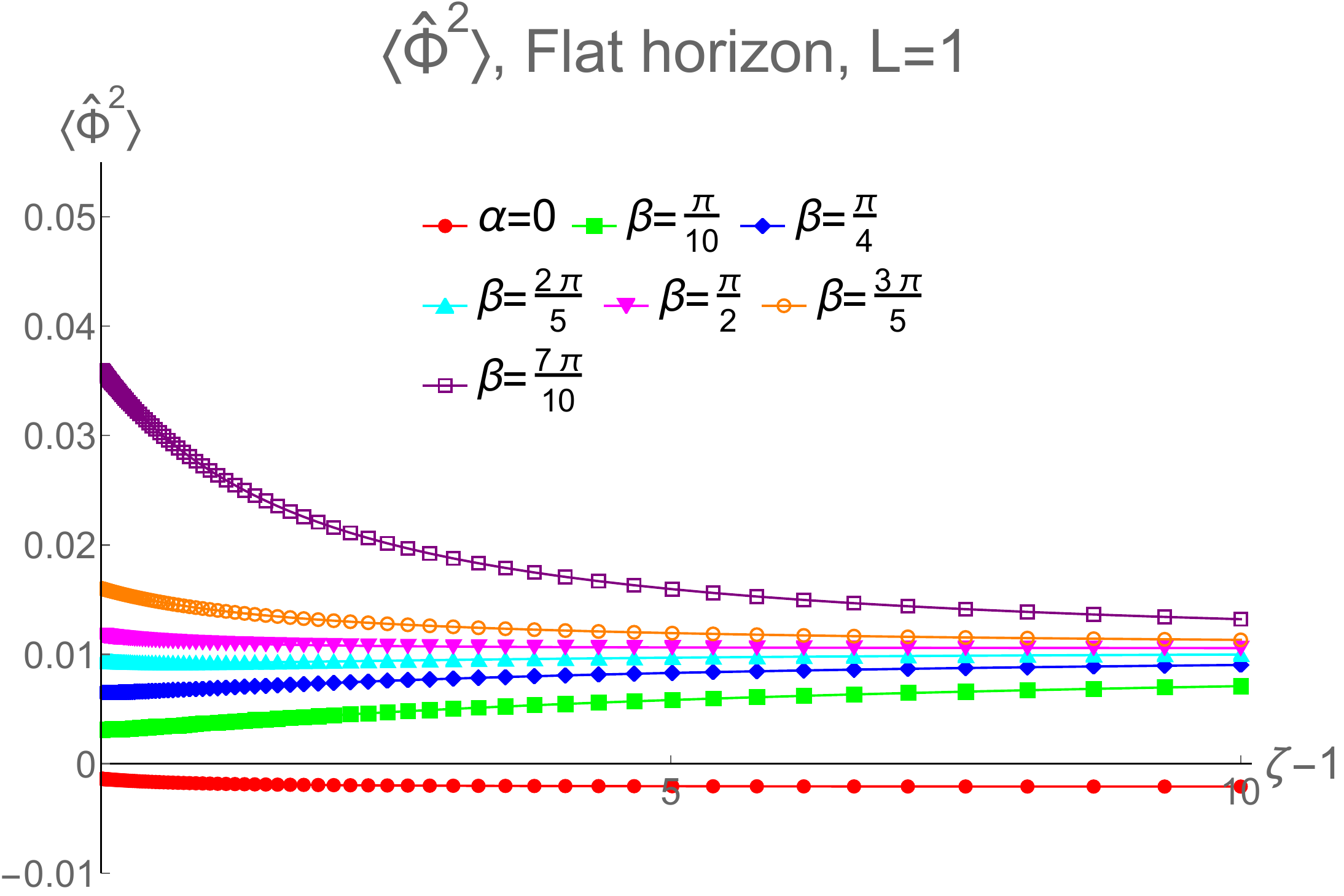}
	\subcaption{$k=0$, $M=4$, $\beta_{{\rm {crit}}}\simeq\frac{71\pi}{100}$}
\end{subfigure}\vspace{10pt}
		\begin{subfigure}[b]{.45\textwidth}
		\centering\includegraphics[width=7cm]{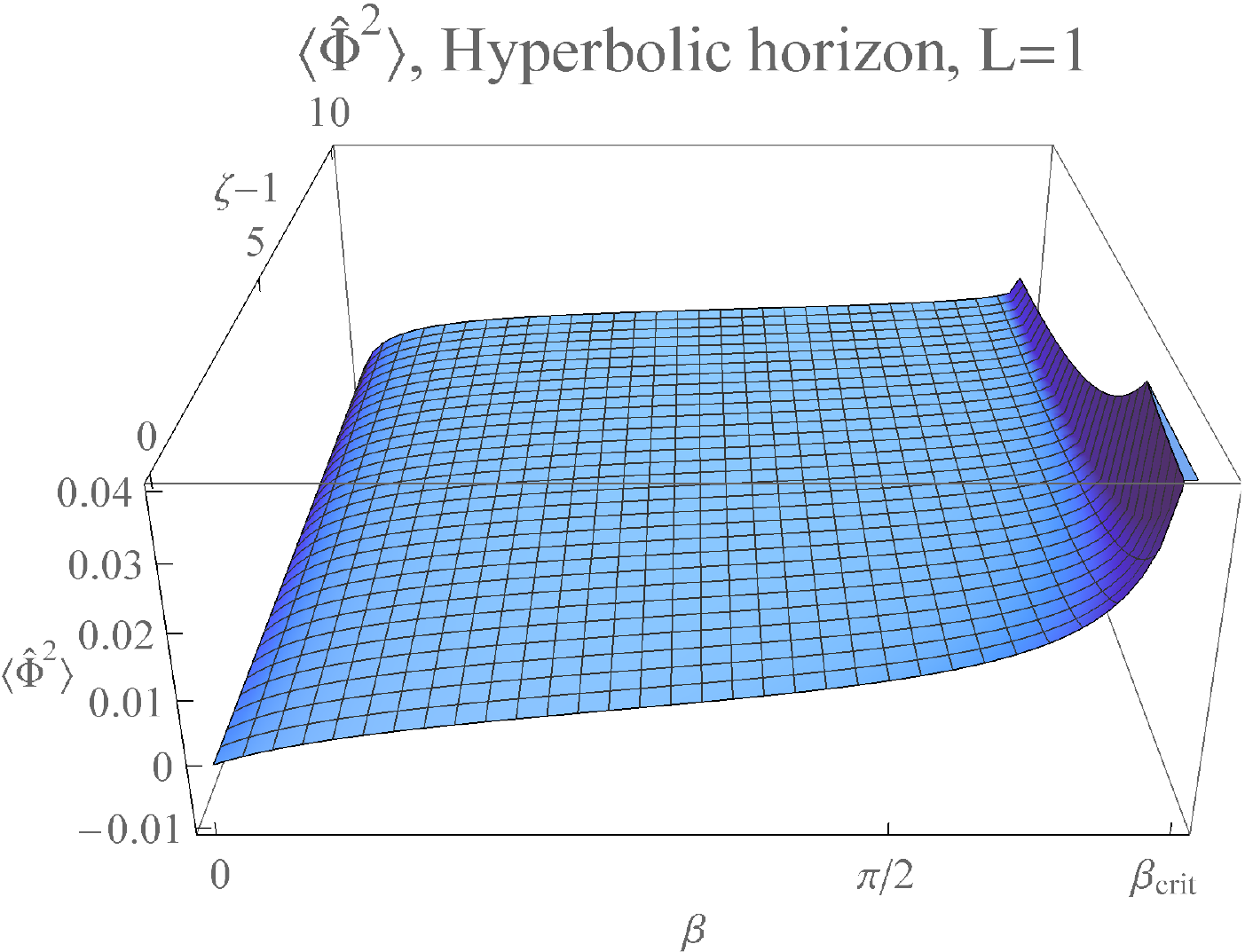}
		\subcaption{$k=-1$, $M=3$, $\beta_{{\rm {crit}}}=2.2411$}
	\end{subfigure}
	\begin{subfigure}[b]{.45\textwidth}
	\centering\includegraphics[width=7cm]{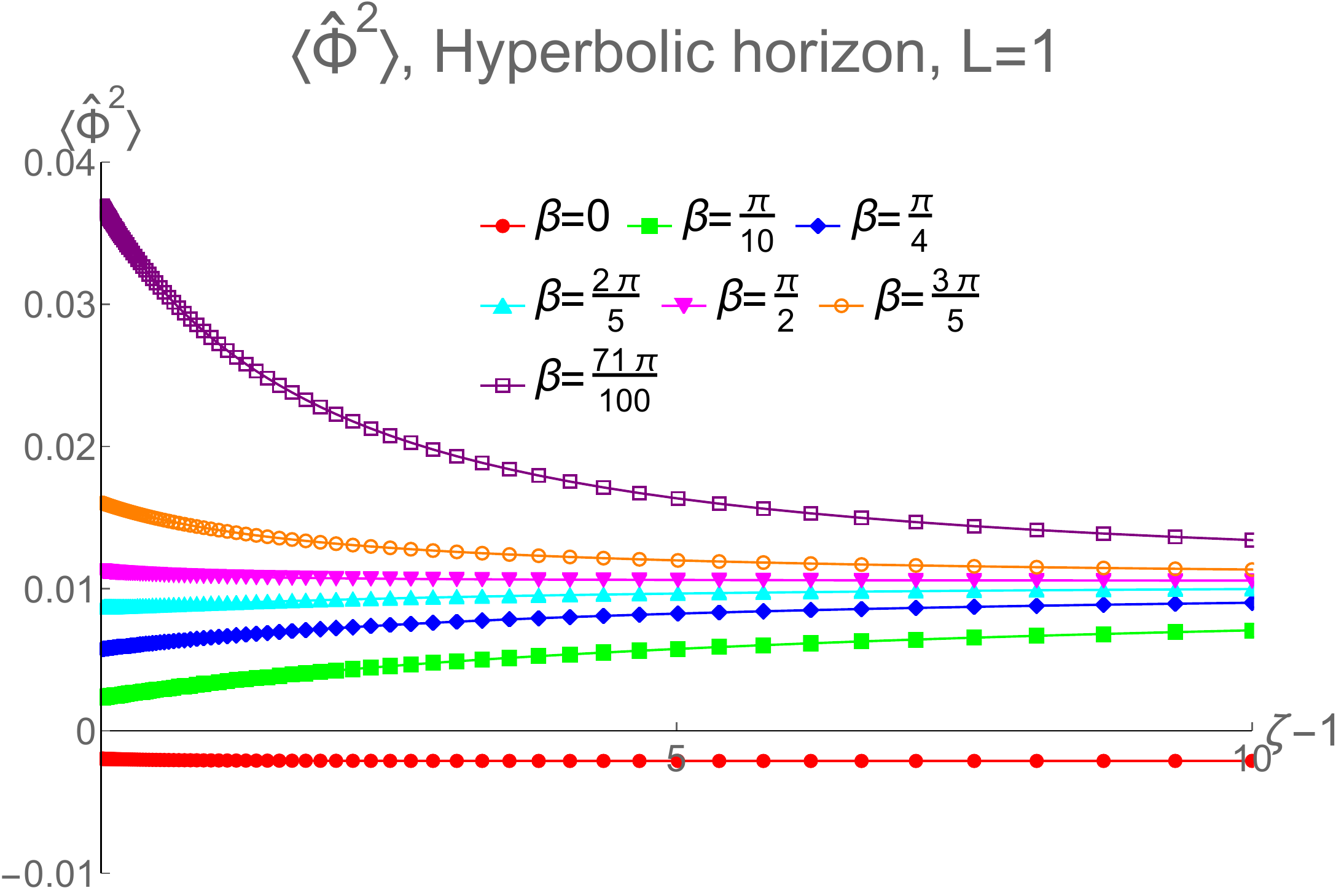}
	\subcaption{$k=-1$, $M=3$, $\beta_{{\rm {crit}}}\simeq\frac{71\pi}{100}$}
\end{subfigure}
	\caption{VP for topological black holes with adS radius of curvature $L=1$ and event horizon radius $r_{h}=2$. \\ Left: surface plots of VP as a function of the dimensionless radial coordinate $\zeta $ (\ref{eq:zetadef}) and parameter $\beta $, for $\beta\in[0,\beta_{{\rm {crit}}})$. Right: line plots of VP as a function of $\zeta $ for a selection of values of $\beta$.}
	\label{fig:L1rh2allbeta}
\end{figure*}

The VP for these black holes with Robin boundary conditions is shown in Fig.~\ref{fig:L1rh2allbeta}. 
In the left-hand plots, we present surface plots of the VP as a function of the parameter $\beta \in [0,\beta _{\rm {crit}})$ and the dimensionless radial coordinate $\zeta $.
The right-hand plots show the profile of the VP as a function of $\zeta $ for a selection of values of $\beta $.
The top plots are for spherical black holes with $k=1$; the middle plots for planar black holes with $k=0$ and the bottom plots for hyperbolic black holes with $k=-1$.
In all cases we see that the value of the VP on the event horizon of the black hole increases as $\beta $ increases, and diverges as $\beta \rightarrow \beta _{\rm {crit}}$.
Similar behaviour was observed in \cite{Morley:2020ayr}, where the value of the VP at the origin in pure adS also increases as the Robin parameter $\alpha $ increases, again diverging as the critical value was approached.
This divergence as $\beta \rightarrow \beta _{\rm {crit}}$ indicates a break-down in the semiclassical approximation, as quantum perturbations of the black hole are no longer small. 
This is to be expected since for $\beta > \beta _{\rm {crit}}$ the dynamics of the scalar field becomes classically unstable. 

In the line plots on the right-hand-side of Fig.~\ref{fig:L1rh2allbeta}, it can be seen that for all values of $\beta $ (except $\beta =0$, corresponding to Dirichlet boundary conditions), far from the black hole the VP approaches the same value, and that value equals the VP at infinity for Neumann boundary conditions (magenta curve). 
Again, we observed similar behaviour for both vacuum and thermal expectation values of the VP in pure adS \cite{Morley:2020ayr}. 
We deduce that, as the spacetime boundary is approached, the VP in the Hartle-Hawking state on a topological black hole background approaches its vacuum value on pure adS spacetime. 
In Fig.~\ref{fig:L1rh2allbeta}, it appears that the profiles of the VP for Dirichlet boundary conditions (red curves) are constants, but this is due to the vertical scale used, as in Fig.~\ref{fig:L1rh2DirNeu} it can be seen that the VP is not constant in this case. 
For Dirichlet boundary conditions, the VP is monotonically decreasing as $\zeta $ increases and we move away from the black hole event horizon.
This is also the case for Neumann boundary conditions $\beta = \pi /2$, and for values of $\beta > \pi /2$.
However, we see from Fig.~\ref{fig:L1rh2allbeta} that there is a range of values of $\beta \in (0,\pi /2)$ for which the VP is monotonically increasing and has a maximum on the spacetime boundary. 
Again, this is similar to the behaviour seen in the pure adS scenario \cite{Morley:2020ayr}.

\begin{figure*}
	\centering
	\begin{subfigure}[b]{.45\textwidth}
		\centering\includegraphics[width=7.5cm]{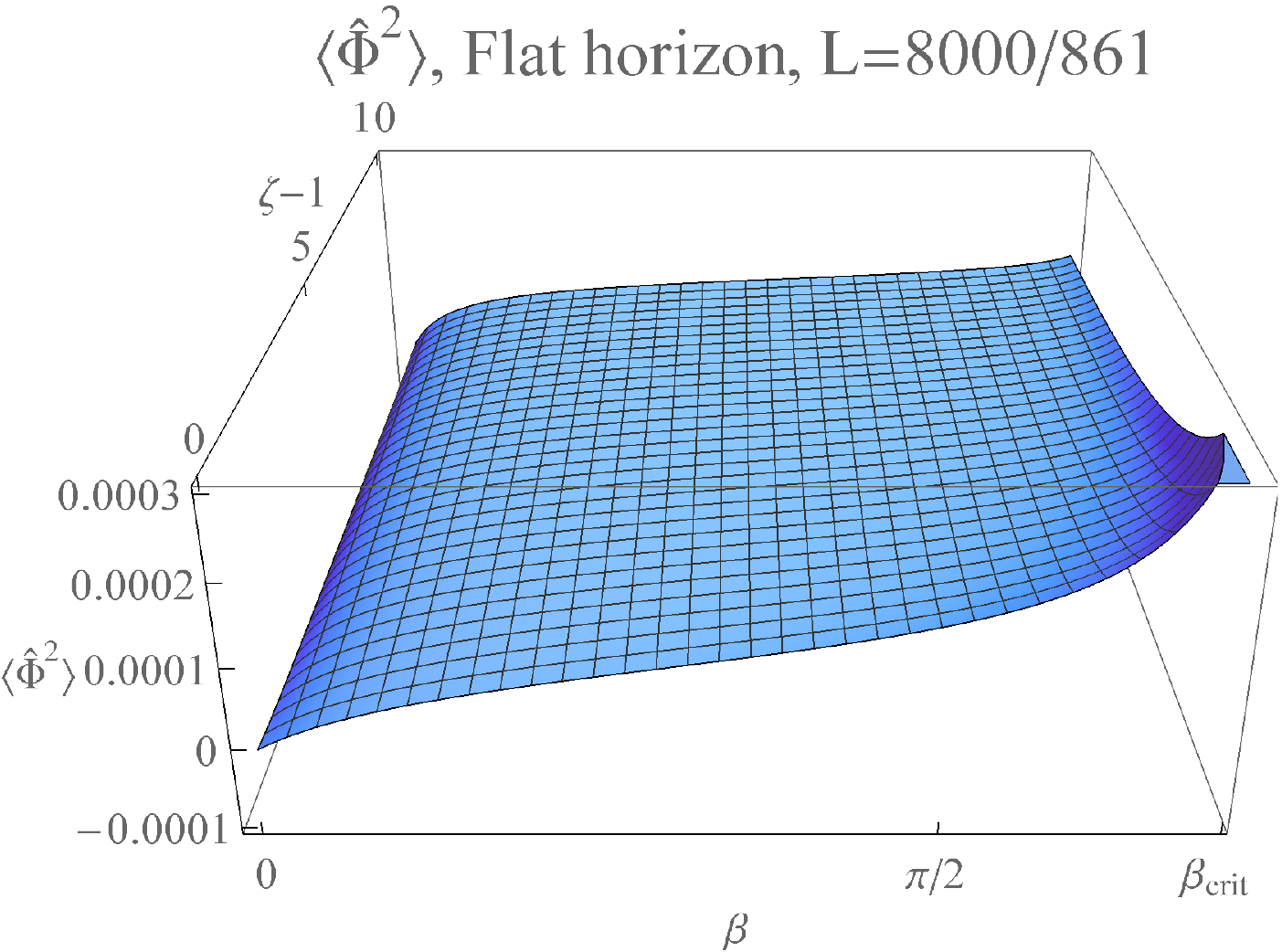}
		\subcaption{$k=0$, $r_{h}=6.8064$, $\beta_{{\rm {crit}}}=2.2300$}
	\end{subfigure}
\begin{subfigure}[b]{.45\textwidth}
	\centering\includegraphics[width=7.5cm]{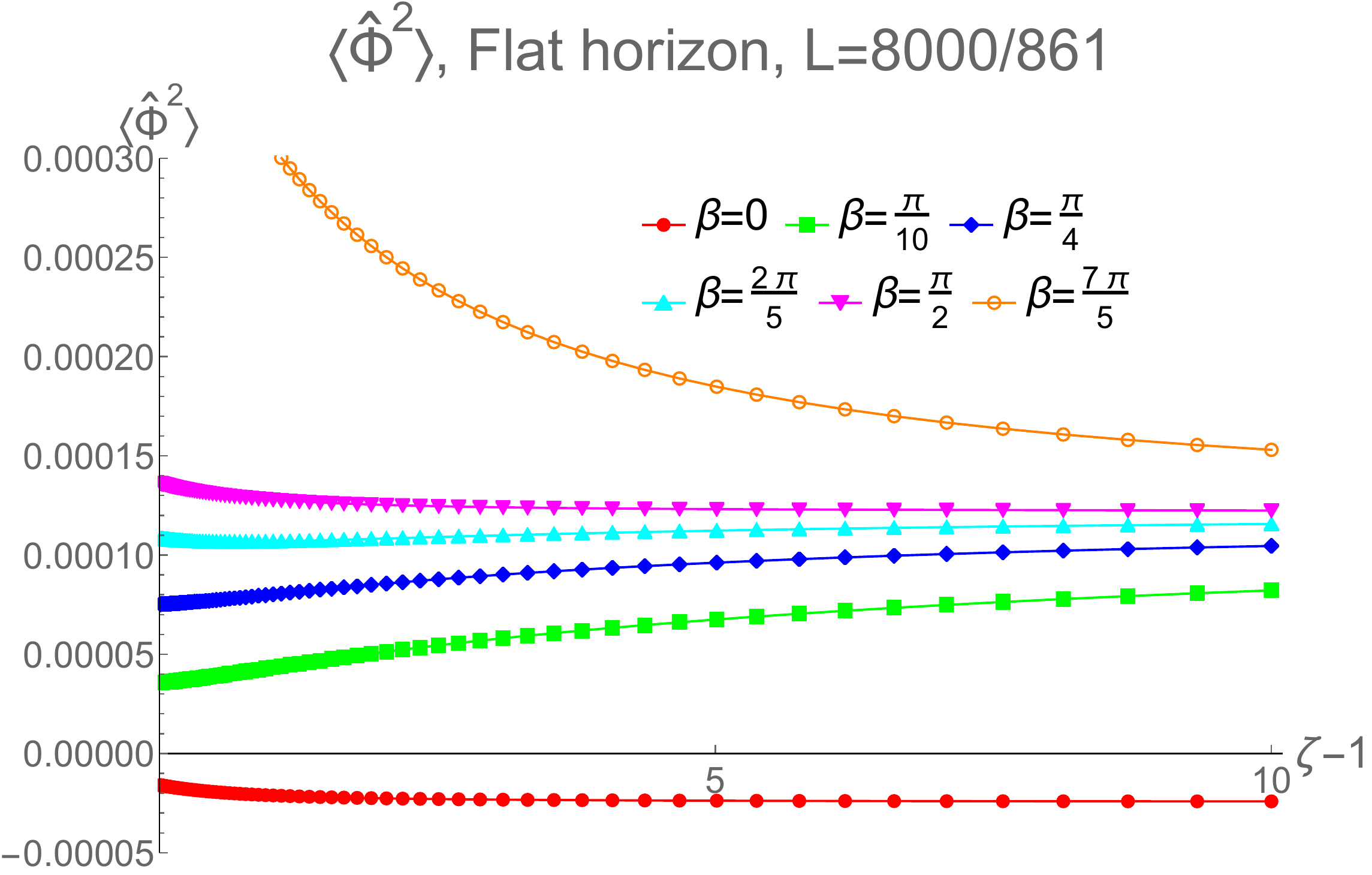}
	\subcaption{$k=0$, $r_{h}=6.8064$, $\beta_{{\rm {crit}}}\simeq\frac{71\pi}{100}$}
\end{subfigure}\vspace{10pt}
	\begin{subfigure}[b]{.45\textwidth}
		\centering\includegraphics[width=7.5cm]{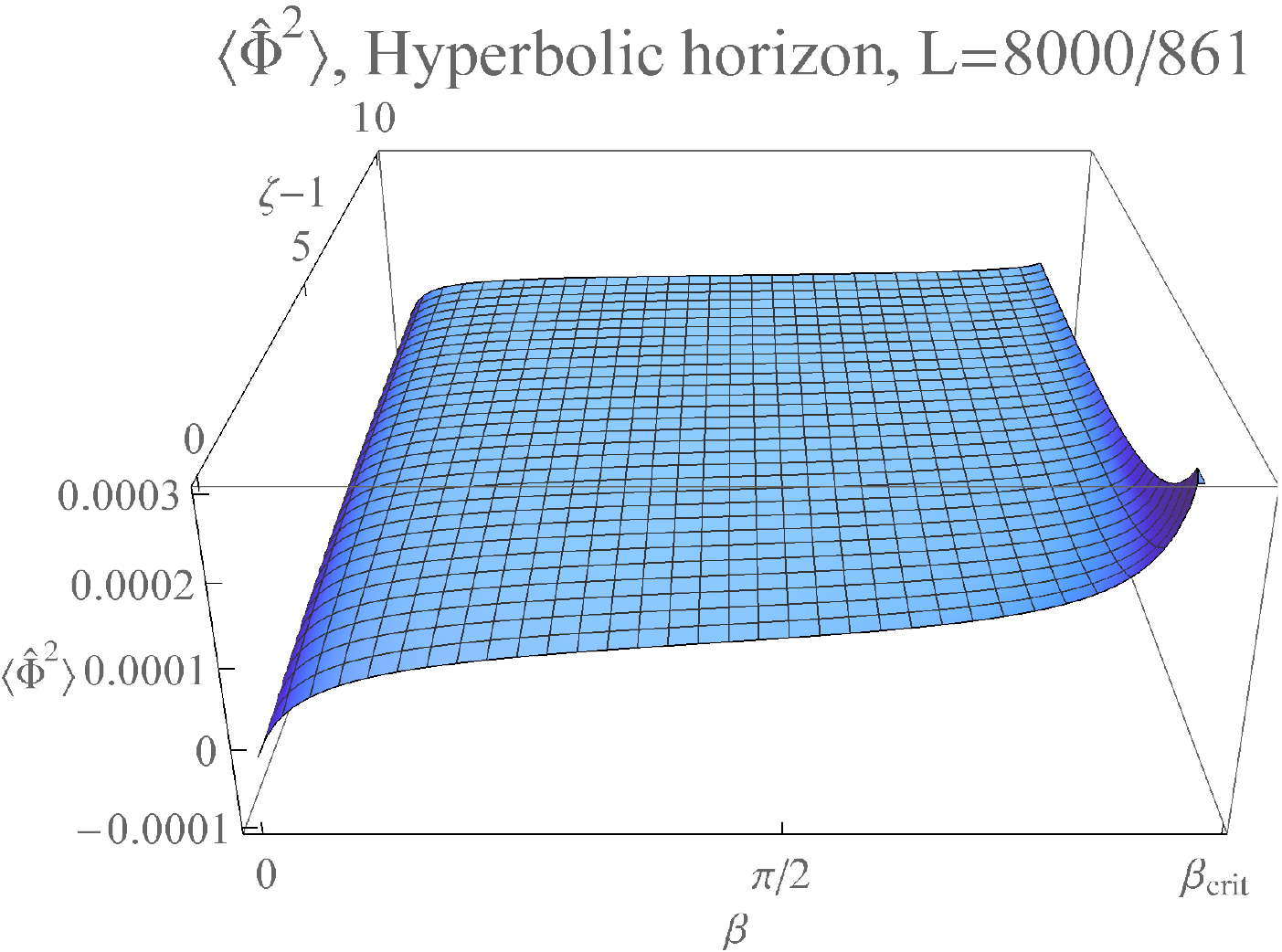}
		\subcaption{$k=-1$, $r_{h}=9.7561$, $\beta_{{\rm {crit}}}=2.9018$}
	\end{subfigure}
	\begin{subfigure}[b]{.45\textwidth}
	\centering\includegraphics[width=7.5cm]{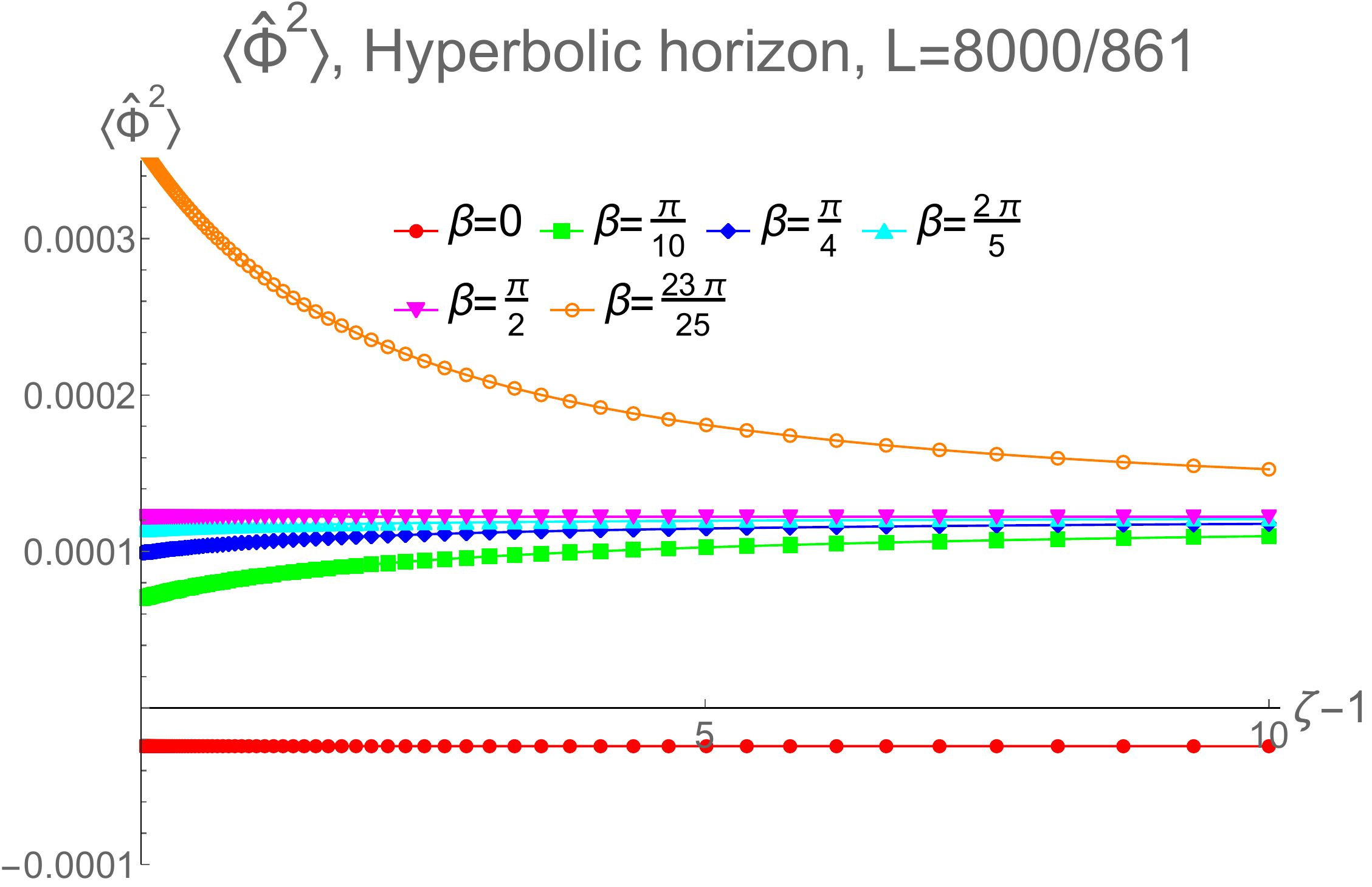}
	\subcaption{$k=-1$, $r_{h}=9.7561$, $\beta_{{\rm {crit}}}\simeq\frac{92\pi}{100}$}
\end{subfigure}
\caption{VP for topological black holes with adS radius of curvature $L=8000/861\approx 9.29$ and surface gravity $\kappa = 37843/320000 \approx 0.12$.
Left: surface plots of VP as a function of the dimensionless radial coordinate $\zeta $ (\ref{eq:zetadef}), and parameter $\beta $, for $\beta \in [0,\beta _{\rm {crit}})$. Right: line plots of VP as a function of $\zeta $ for a selection of values of $\beta $.}
\label{fig:case1}
\end{figure*}

Next, in Fig.~\ref{fig:case1}, we study the VP for low-temperature planar and hyperbolic black holes, the temperature being below the minimum temperature $T_{\rm {min}}$ (\ref{eq:Tmin}) for the existence of spherical black holes. 
The adS radius of curvature is fairly large $L\approx 9.29$ and the surface gravity $\kappa \approx 0.12$. 
The $k=-1$ black holes at this temperature are rather larger ($r_{h}=9.7561$) than the $k=0$ black holes ($r_{h}=6.8064$) having the same adS radius of curvature. 
Since the temperature is very low, the VP has very small values, and is much smaller in magnitude than the VP depicted in Fig.~\ref{fig:L1rh2allbeta}.
The VP has very similar qualitative behaviour to that shown in Fig.~\ref{fig:L1rh2allbeta}. 
In particular, the VP is a monotonically decreasing function of the dimensionless radial coordinate $\zeta $ except for values of $\beta $ in an interval contained in $(0,\pi /2)$. 
On the event horizon, the VP increases as the parameter $\beta $ increases, and far from the black hole, the VP for all values of $\beta $ except $\beta =0$ approaches the vacuum value in pure adS spacetime for Neumann boundary conditions (\ref{eq:adSN}). 

\begin{figure*}
	\centering
	\includegraphics[width=0.45\textwidth]{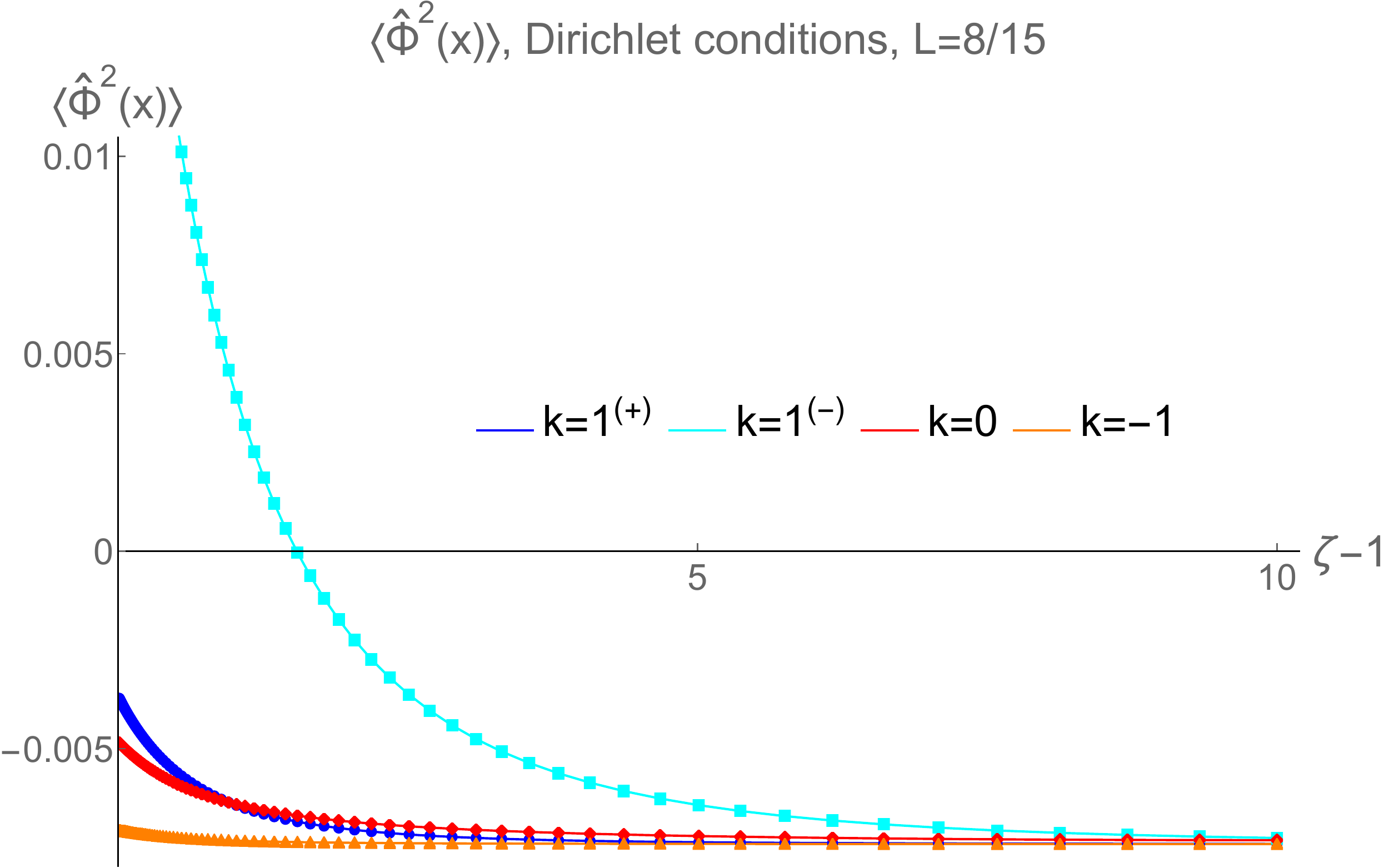}
	\includegraphics[width=0.45\textwidth]{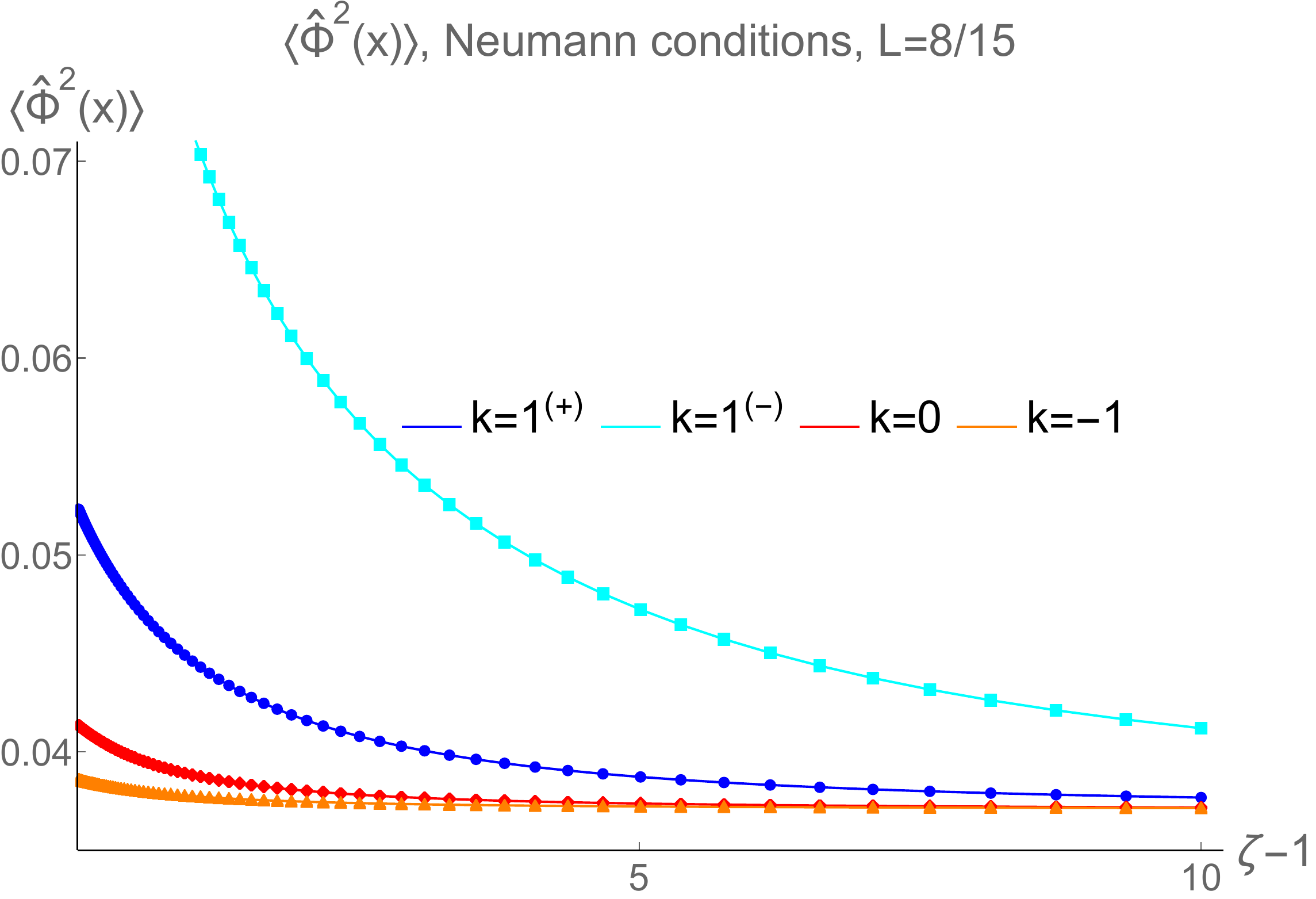}
	\caption{VP for topological black holes with adS radius of curvature $L=8/15\approx 0.53$ and surface gravity $\kappa = 115/32 \approx 3.59$, with Dirichlet (left) and Neumann (right) boundary conditions applied.}
	\label{fig:L8_15DirNeu}
\end{figure*}

\begin{figure*}
	\centering
	\begin{subfigure}[b]{.4\textwidth}
		\centering\includegraphics[width=6.5cm]{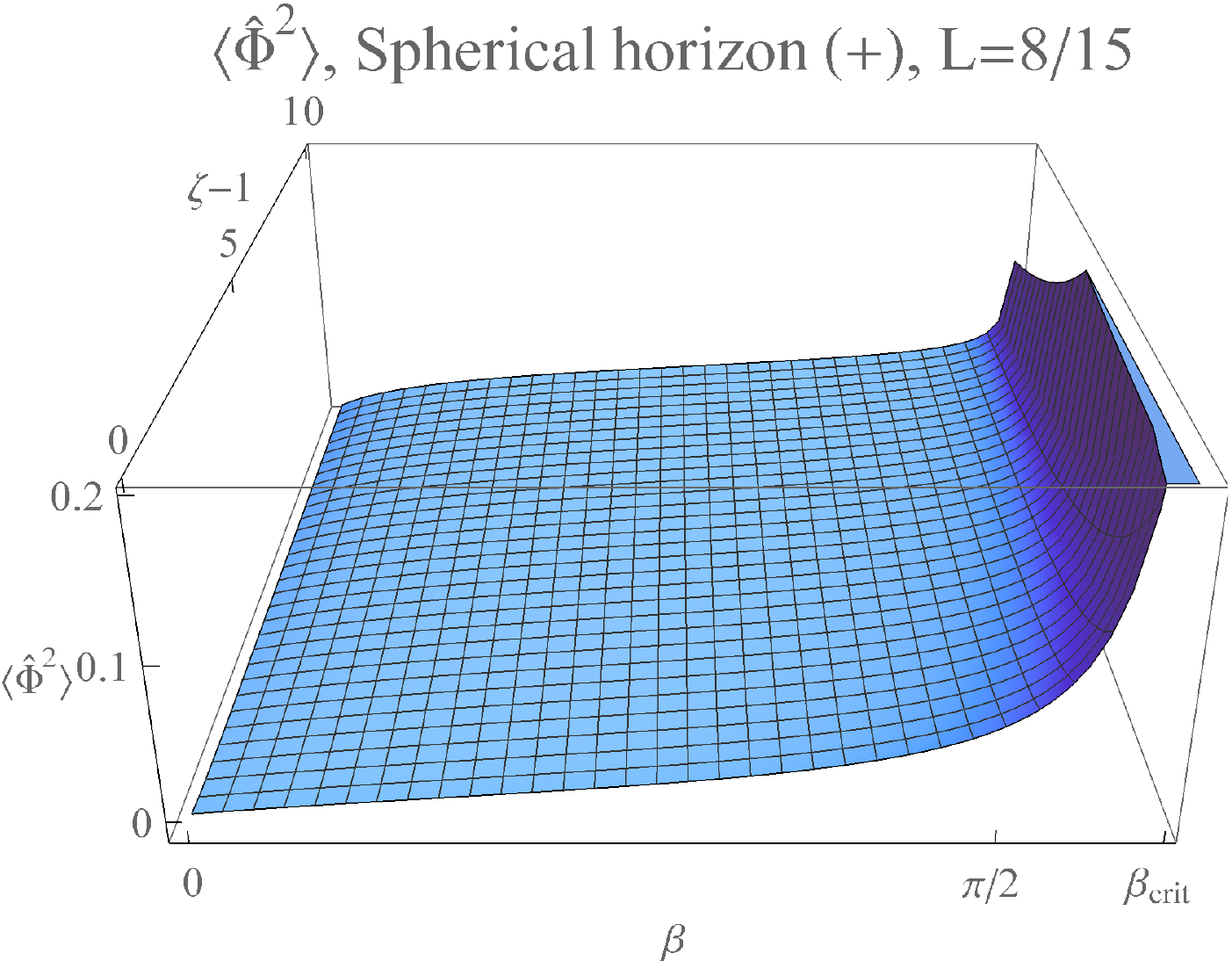}
		\subcaption{$k=1^{(+)}$, $r_{h}=0.1948$, $\beta_{{\rm {crit}}}=1.8833$}
	\end{subfigure}\hspace{2cm}
	\begin{subfigure}[b]{.4\textwidth}
		\centering\includegraphics[width=6.5cm]{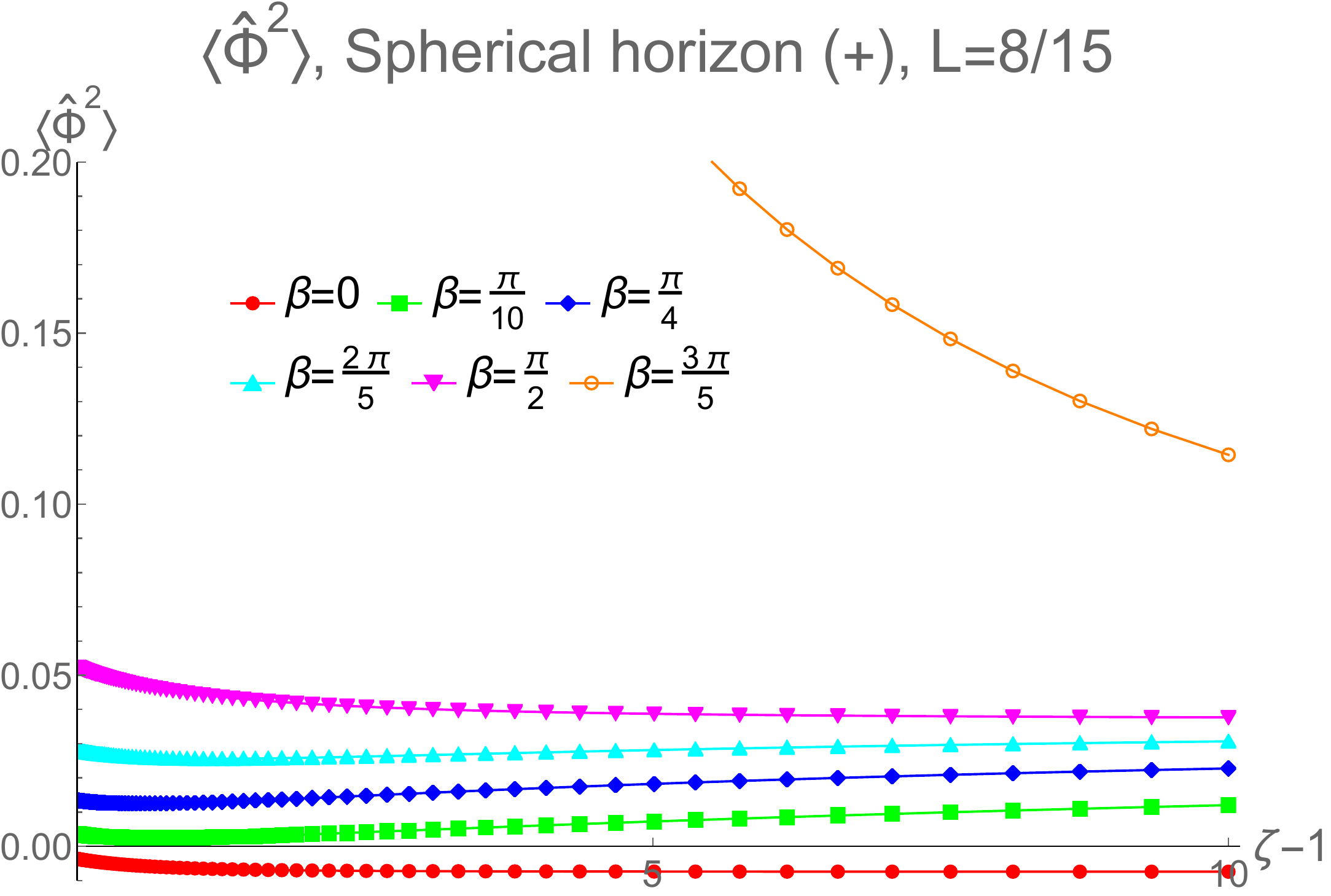}
		\subcaption{$k=1^{(+)}$, $r_{h}=0.1948$, $\beta_{{\rm {crit}}}\simeq\frac{60\pi}{100}$}
	\end{subfigure}\vspace{5pt}
	\begin{subfigure}[b]{.4\textwidth}
		\centering\includegraphics[width=6.5cm]{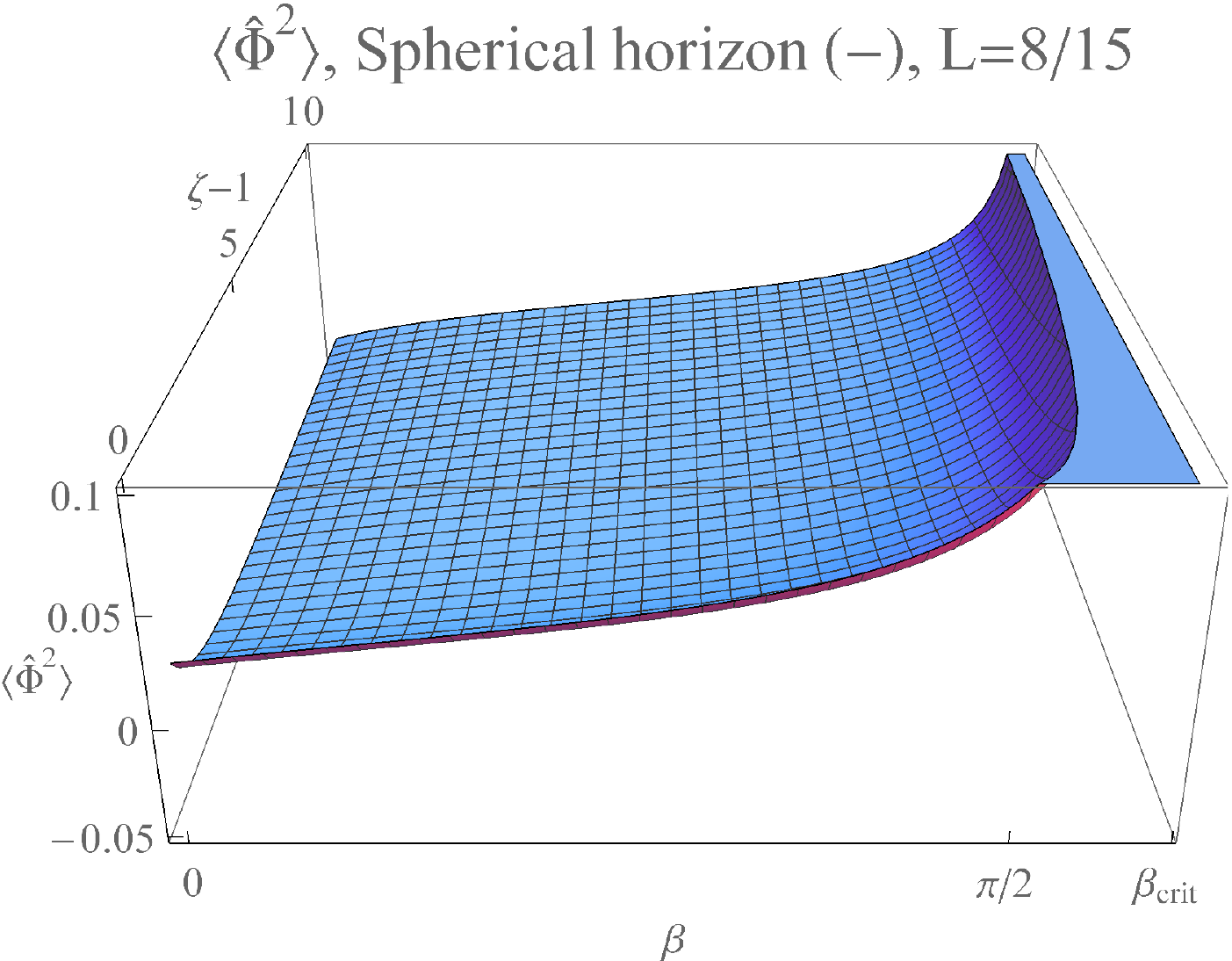}
		\subcaption{$k=1^{(-)}$, $r_{h}=0.4866$, $\beta_{{\rm {crit}}}=1.8998$}
	\end{subfigure}\hspace{2cm}
	\begin{subfigure}[b]{.4\textwidth}
		\centering\includegraphics[width=6.5cm]{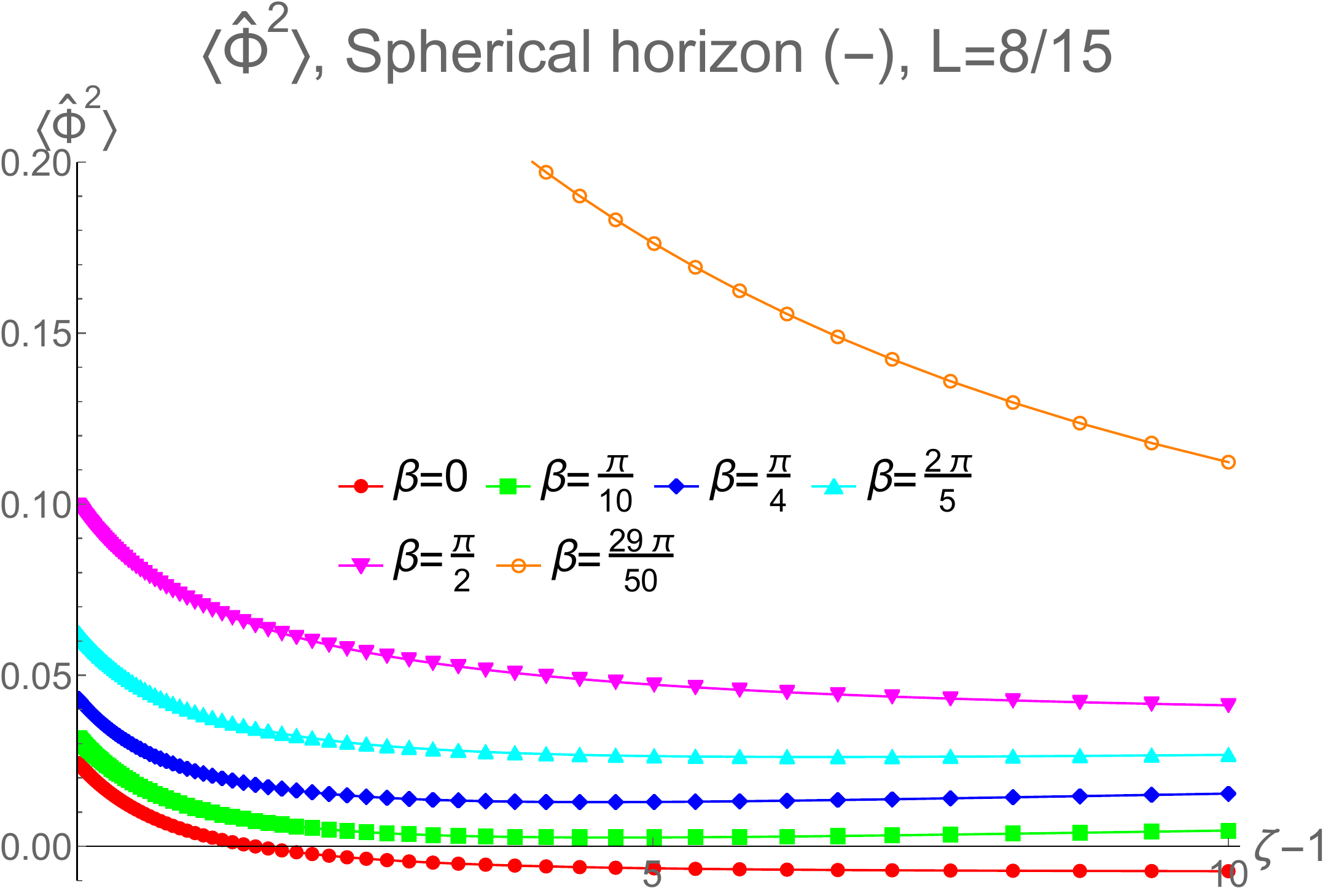}
		\subcaption{$k=1^{(-)}$, $r_{h}=0.4866$, $\beta_{{\rm {crit}}}\simeq\frac{60\pi}{100}$}
	\end{subfigure}\vspace{5pt}
	\begin{subfigure}[b]{.4\textwidth}
		\centering\includegraphics[width=6.5cm]{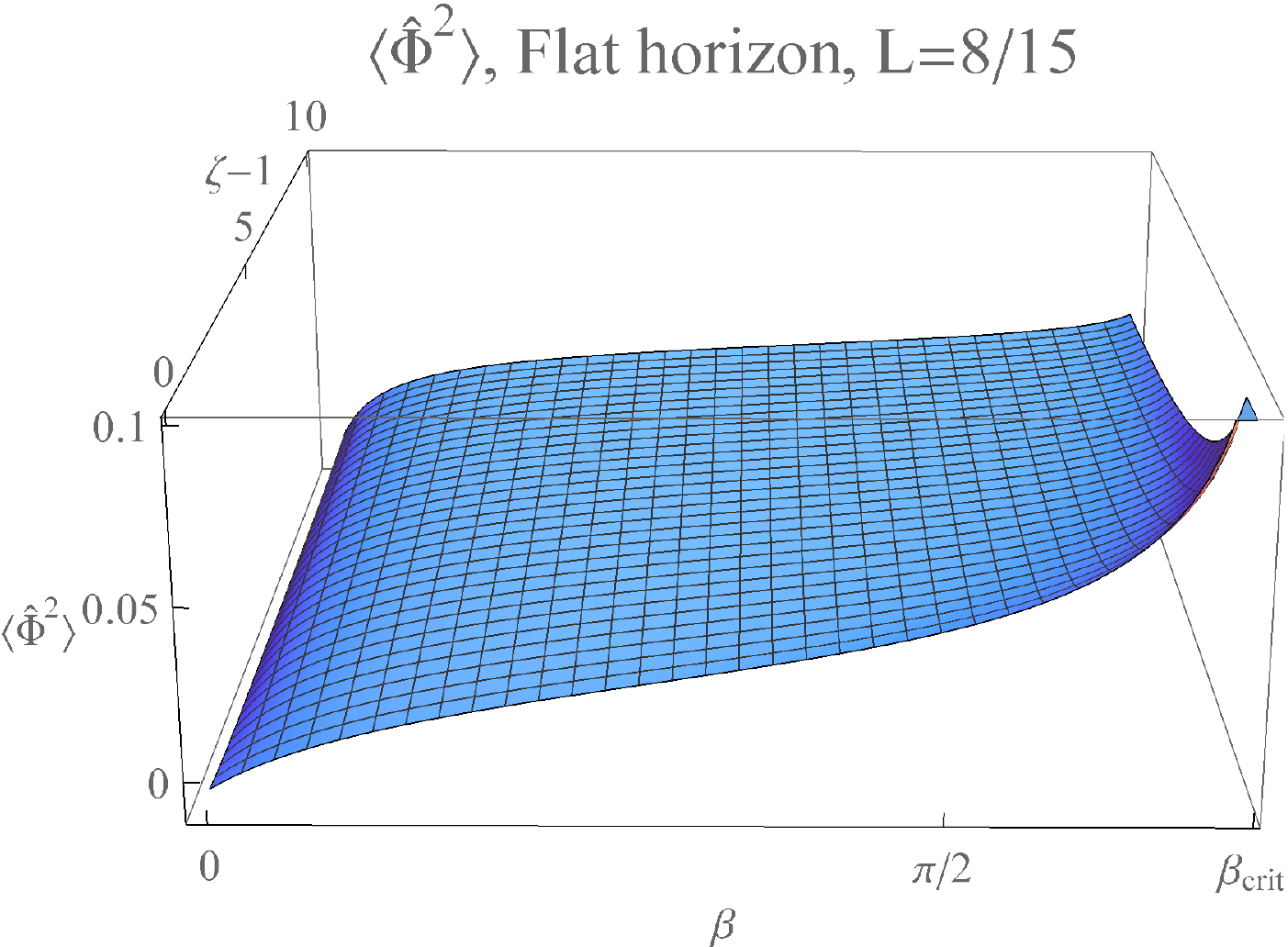}
		\subcaption{$k=0$, $r_{h}=0.6815$, $\beta_{{\rm {crit}}}=2.2301$}
	\end{subfigure}\hspace{2cm}
	\begin{subfigure}[b]{.4\textwidth}
		\centering\includegraphics[width=6.5cm]{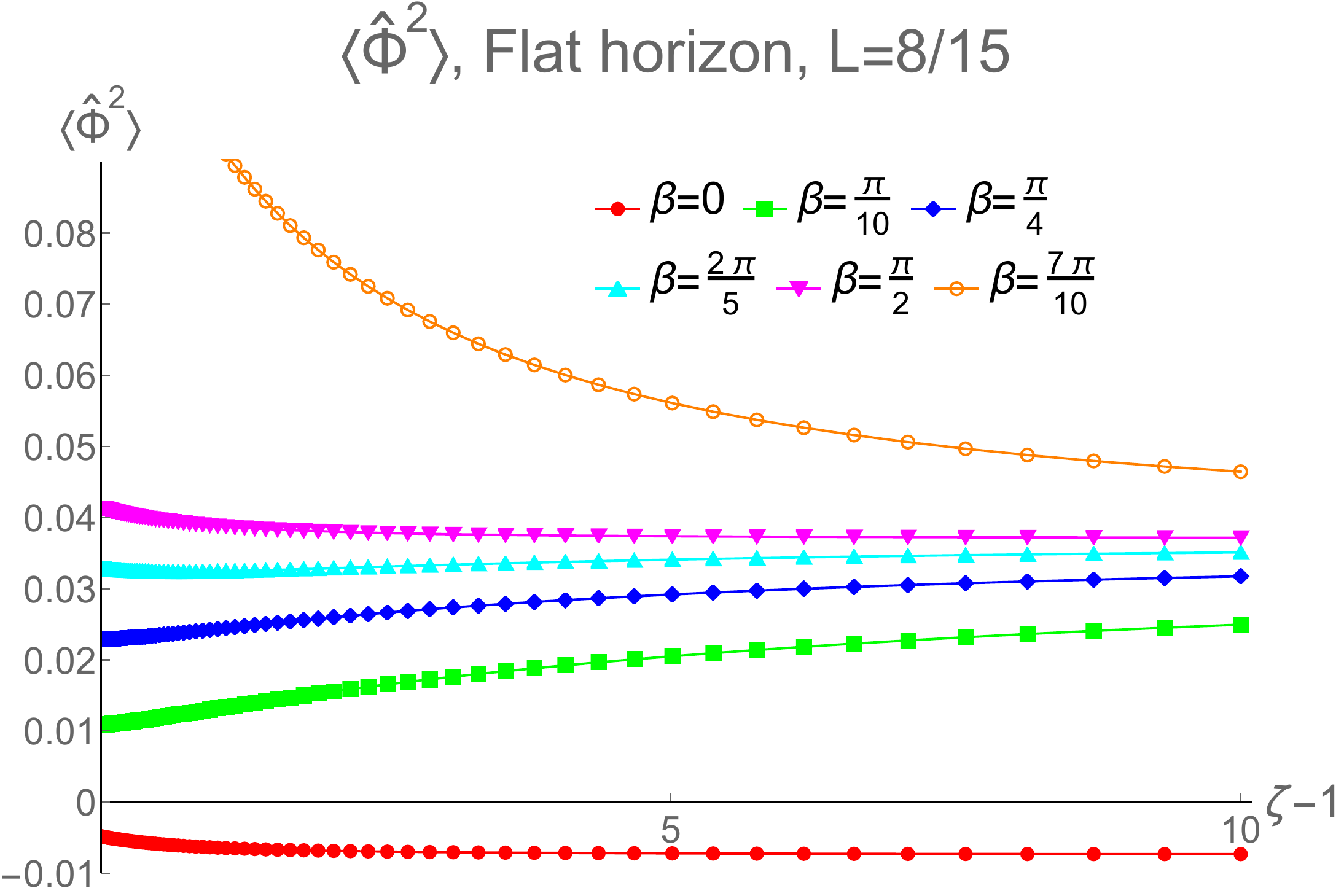}
		\subcaption{$k=0$, $r_{h}=0.6815$, $\beta_{{\rm {crit}}}\simeq\frac{71\pi}{100}$}
	\end{subfigure}\vspace{5pt}
	\begin{subfigure}[b]{.4\textwidth}
		\centering\includegraphics[width=6.5cm]{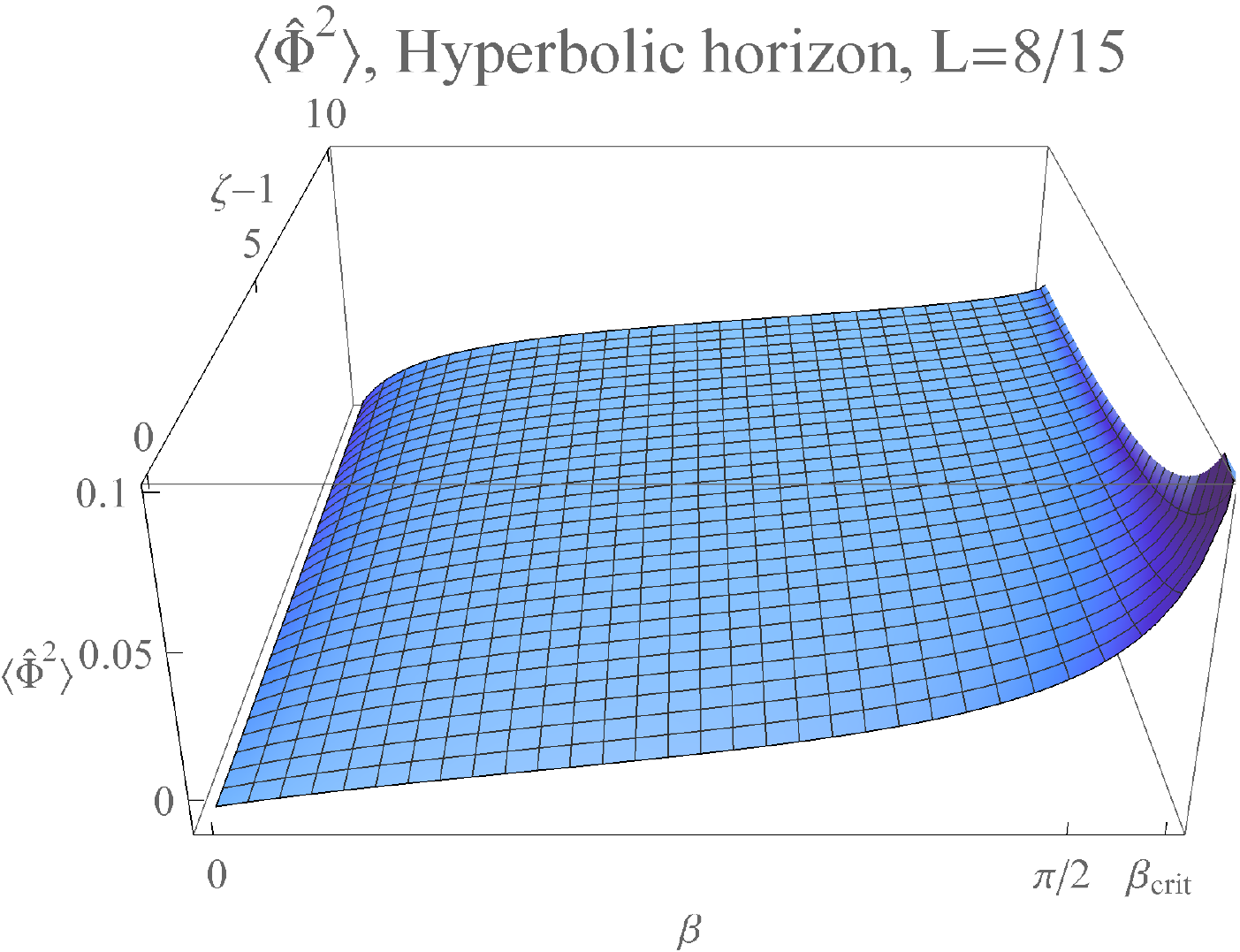}
		\subcaption{$k=-1$, $r_{h}=0.8$, $\beta_{{\rm {crit}}}=1.8829$}
	\end{subfigure}\hspace{2cm}
\begin{subfigure}[b]{.4\textwidth}
	\centering\includegraphics[width=6.5cm]{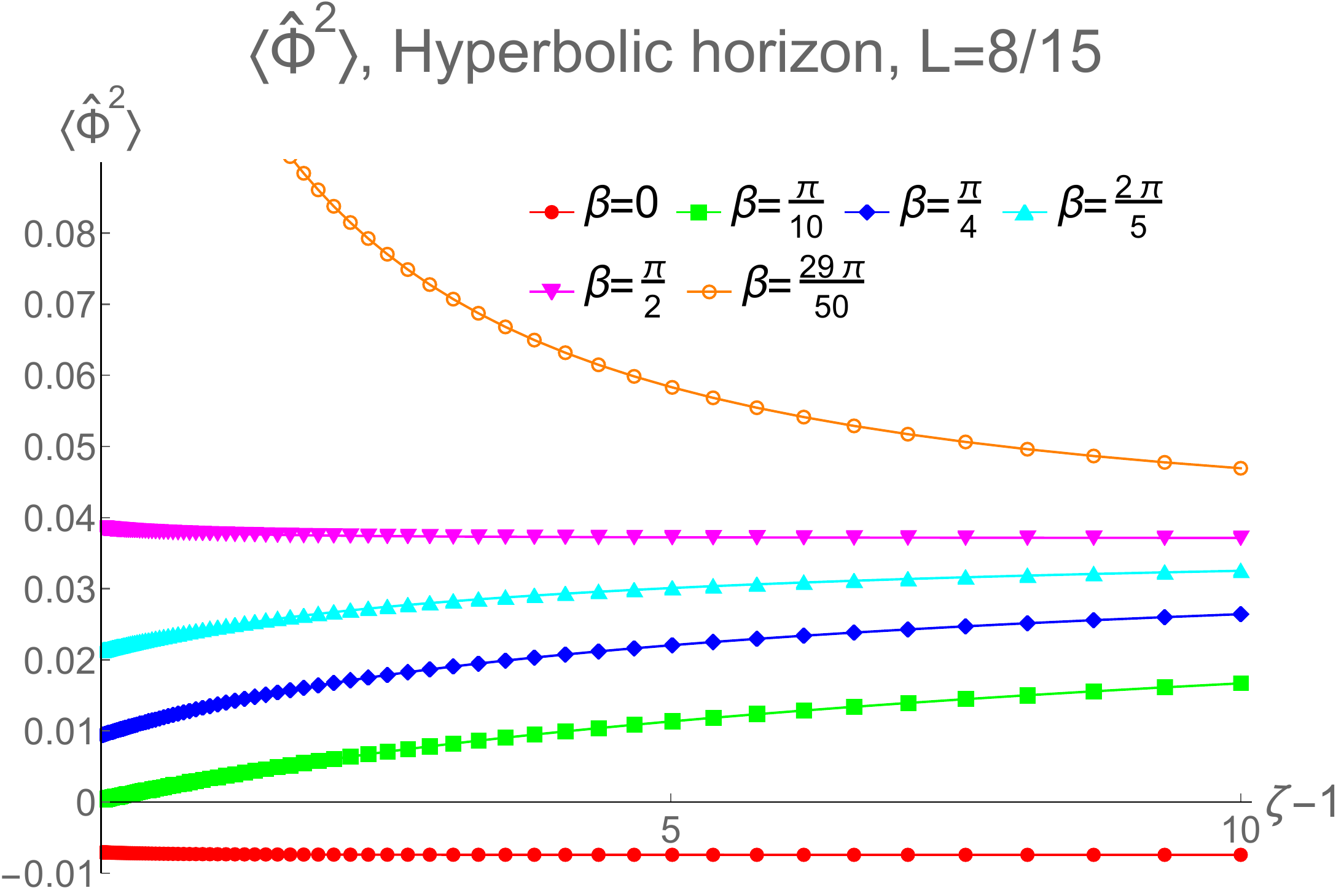}
	\subcaption{$k=-1$, $r_{h}=0.8$, $\beta_{{\rm {crit}}}\simeq\frac{60\pi}{100}$}
\end{subfigure}
\caption{VP for topological black holes with adS radius of curvature $L=8/15\approx 0.53$ and surface gravity $\kappa = 115/32 \approx 3.59$.
	Left: surface plots of VP as a function of the dimensionless radial coordinate $\zeta $ (\ref{eq:zetadef}), and parameter $\beta $, for $\beta \in [0,\beta _{\rm {crit}})$. Right: line plots of VP as a function of $\zeta $ for a selection of values of $\beta $.}
\label{fig:case2}
\end{figure*}

\begin{figure*}
	\centering
	\begin{subfigure}[b]{.45\textwidth}
		\centering\includegraphics[width=7.5cm]{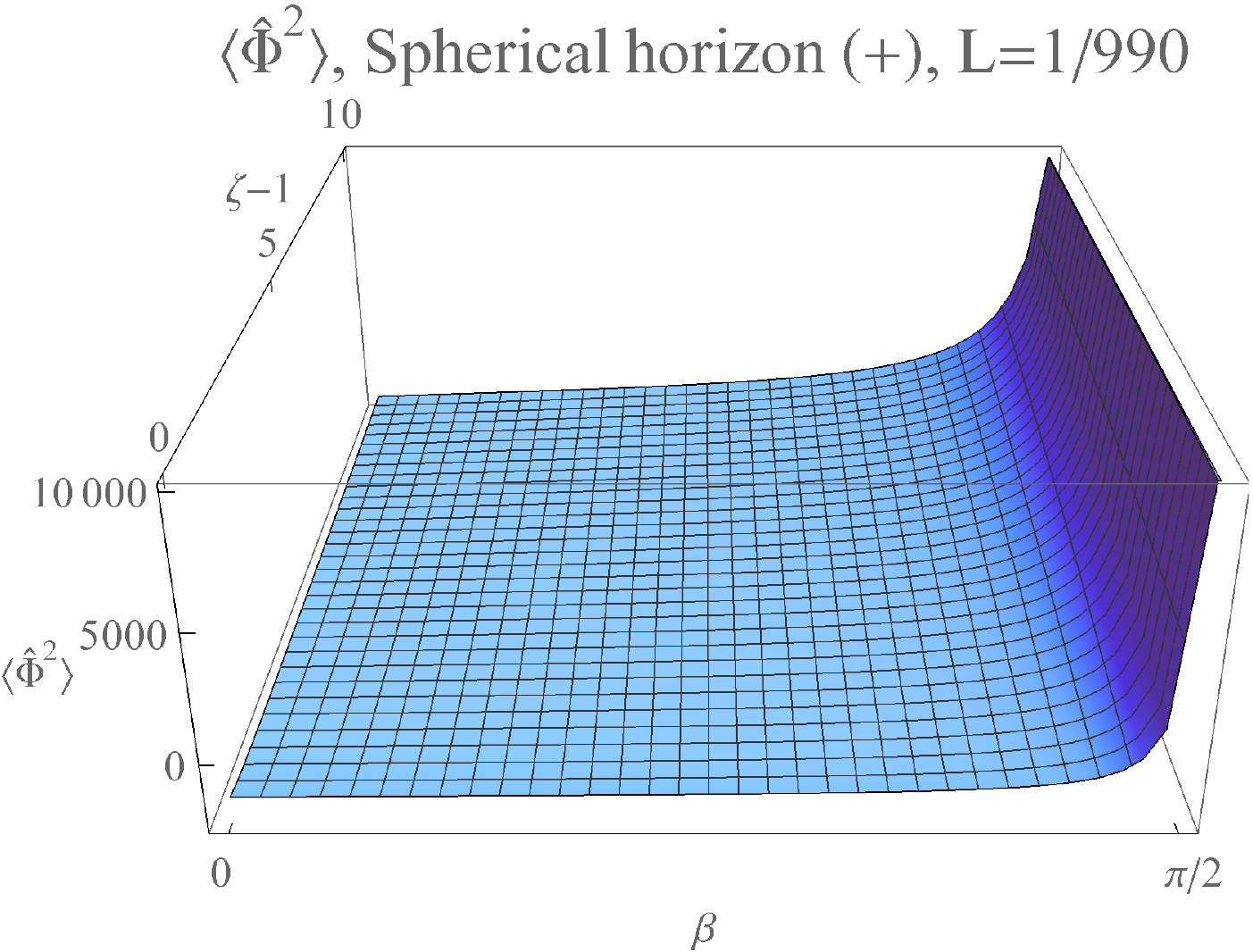}
		\subcaption{$k=1^{(+)}$, $r_{h}=0.0101$, $\beta_{{\rm {crit}}}=1.5747$}
	\end{subfigure}
		\begin{subfigure}[b]{.45\textwidth}
		\centering\includegraphics[width=7.5cm]{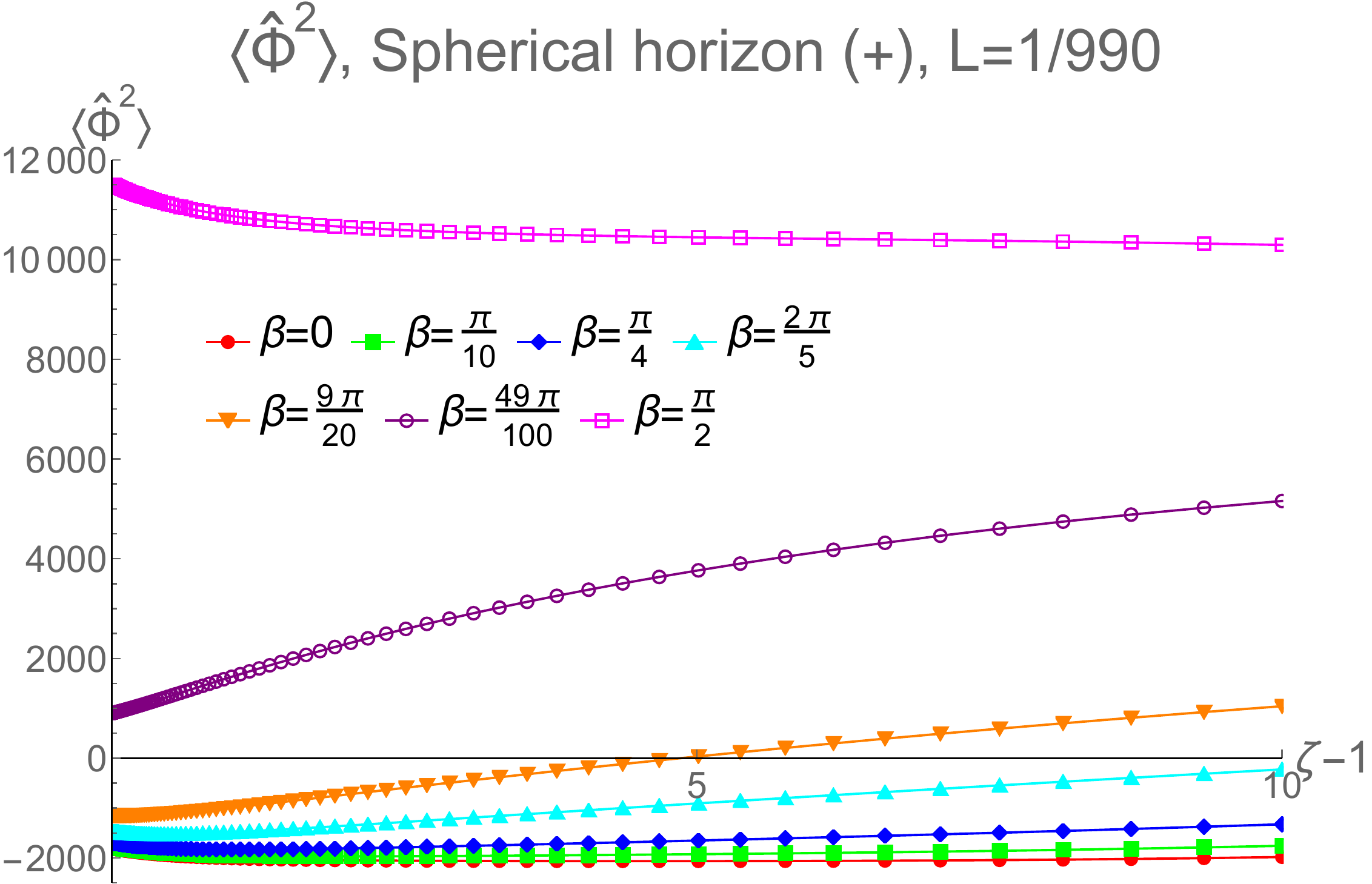}
		\subcaption{$k=1^{(+)}$, $r_{h}=0.0101$, $\beta_{{\rm {crit}}}\simeq\frac{50\pi}{100}$}
	\end{subfigure}\vspace{10pt}	
	\begin{subfigure}[b]{.45\textwidth}
		\centering\includegraphics[width=7.5cm]{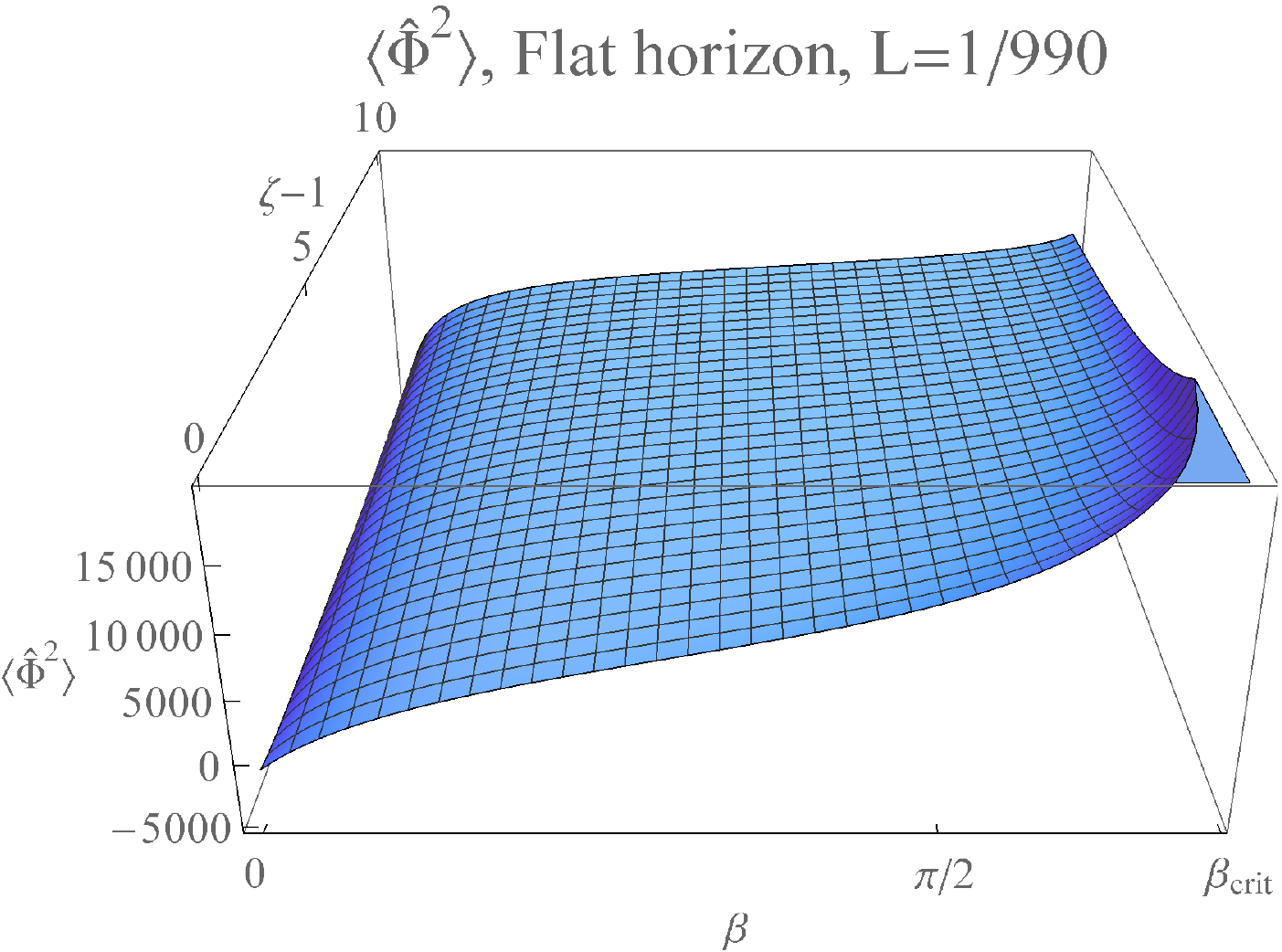}
		\subcaption{$k=0$, $r_{h}=0.0101$, $\beta_{{\rm {crit}}}=2.2302$}
	\end{subfigure}
	\begin{subfigure}[b]{.45\textwidth}
		\centering\includegraphics[width=7.5cm]{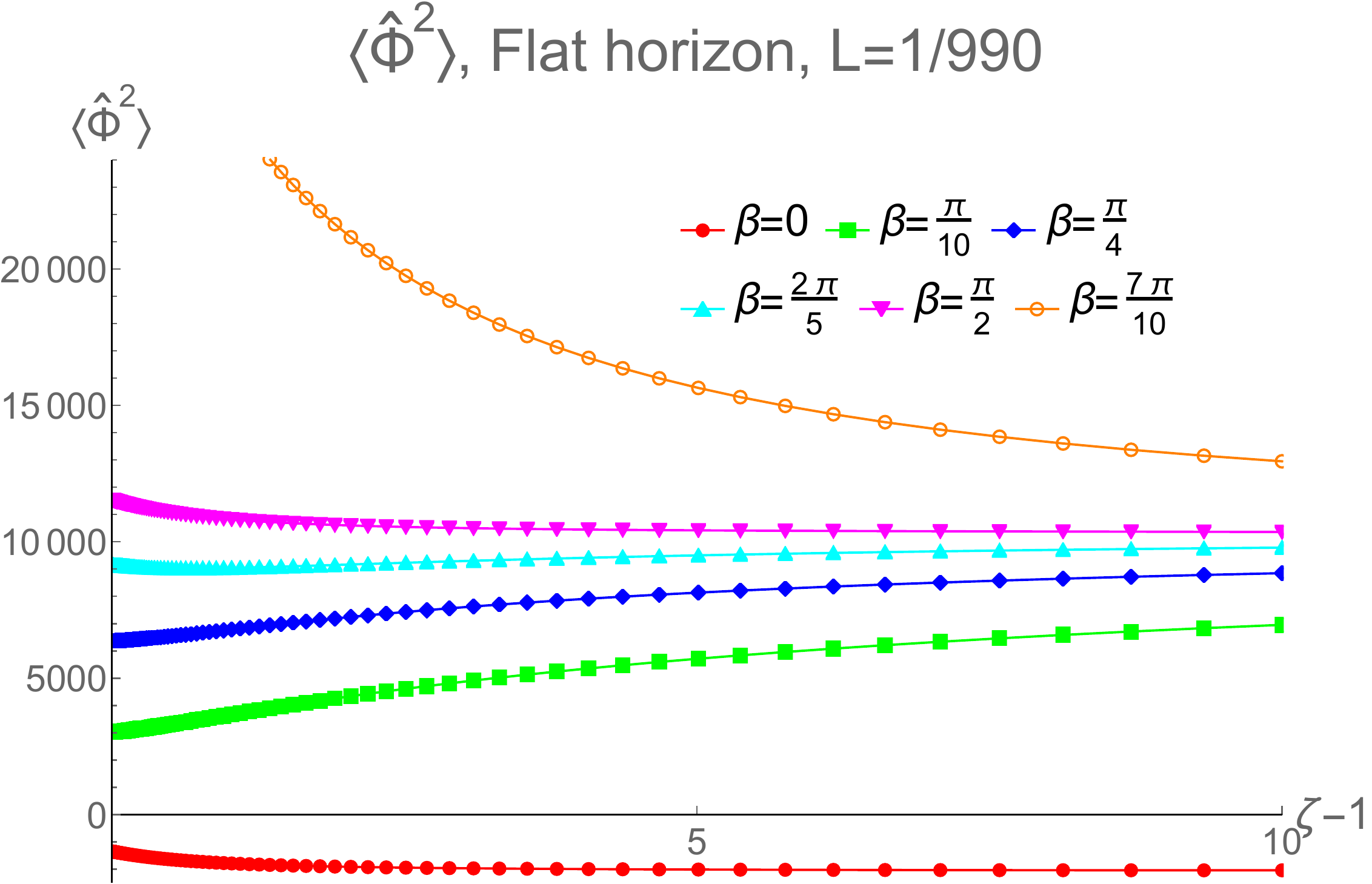}
		\subcaption{$k=0$, $r_{h}=0.0101$, $\beta_{{\rm {crit}}}\simeq\frac{71\pi}{100}$}
	\end{subfigure}\vspace{10pt}
	\begin{subfigure}[b]{.45\textwidth}
		\centering\includegraphics[width=7.5cm]{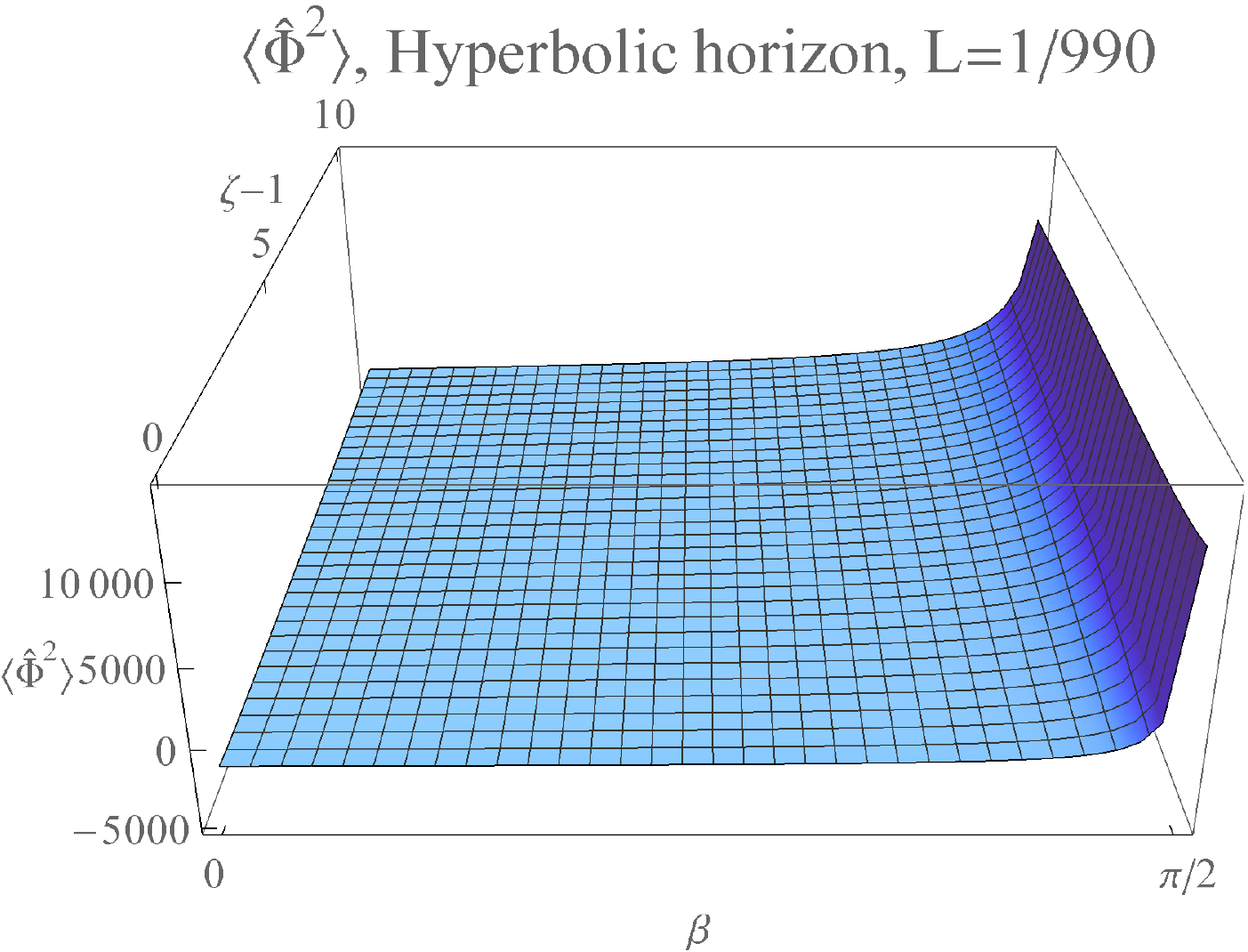}
		\subcaption{$k=-1$, $r_{h}=0.0101$, $\beta_{{\rm {crit}}}=1.5747$}
	\end{subfigure}
	\begin{subfigure}[b]{.45\textwidth}
	\centering\includegraphics[width=7cm]{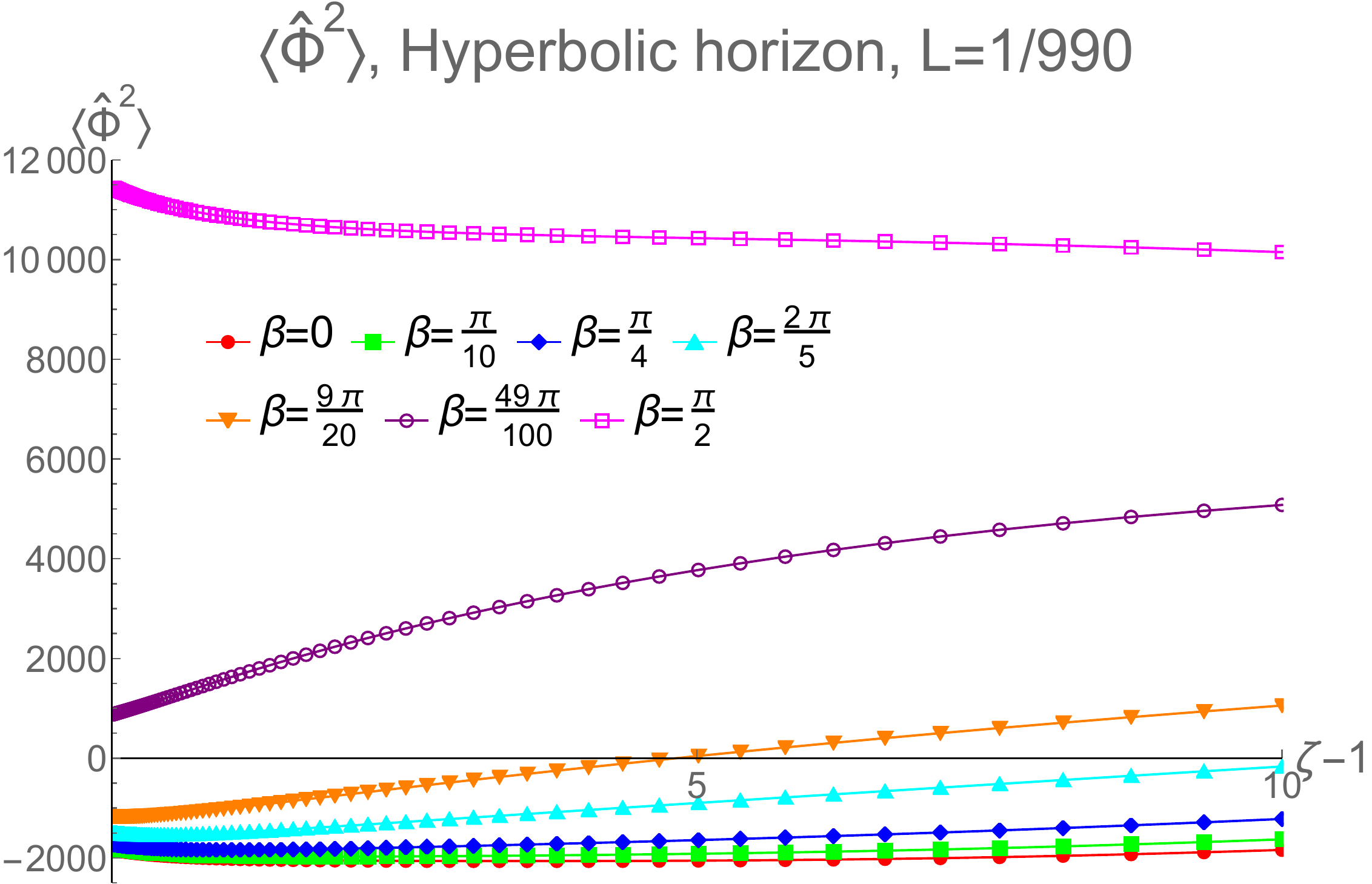}
	\subcaption{$k=-1$, $r_{h}=0.0101$, $\beta_{{\rm {crit}}}\simeq\frac{50\pi}{100}$}
\end{subfigure}
\caption{VP for topological black holes with adS radius of curvature $L=1/990\approx 0.001$ and surface gravity $\kappa = 29601/2\approx 14801$.
	Left: surface plots of VP as a function of the dimensionless radial coordinate $\zeta $ (\ref{eq:zetadef}), and parameter $\beta $, for $\beta \in [0,\beta _{\rm {crit}})$. Right: line plots of VP as a function of $\zeta $ for a selection of values of $\beta $.}
\label{fig:case3}
\end{figure*}

In our remaining plots we increase the temperature of the black hole so that, in addition to the $k=0$ and $k=-1$ black holes, there are also two black holes with spherical horizons, one ($k=1^{(+)}$) which is larger and thermodynamically stable, and a smaller ($k=1^{(-)}$), thermodynamically unstable, black hole. 
In Figs.~\ref{fig:L8_15DirNeu}--\ref{fig:case2} the black hole temperature is fairly close to the minimum for the existence of spherical black holes, while in Fig.~\ref{fig:case3} we consider black holes having a large temperature. 

First we consider black holes with adS radius of curvature $L\approx 0.53$ and $\kappa \approx 3.59$.
In Fig.~\ref{fig:L8_15DirNeu}, we show the VP for Dirichlet (left) and Neumann (right) boundary conditions, comparing the results for the black holes with different $k$. 
In both cases the VP decreases monotonically from the event horizon to its value (\ref{eq:adS}) at the spacetime boundary.  
The VP for the thermodynamically unstable spherical black hole ($k=1^{(-)}$, light blue curve) has significantly larger value on the horizon than for the thermodynamically stable spherical black hole ($k=1^{(+)}$). 
Unlike the situation in Fig.~\ref{fig:L1rh2DirNeu}, here the order of the curves is the same for both Dirichlet and Neumann boundary conditions.
The $k=1^{(-)}$ curve always has the largest VP on the horizon, followed by the $k=1^{(+)}$ curve, then the planar $k=0$ black hole and finally the hyperbolic black hole with $k=-1$ always has the smallest VP on the horizon.

Fig.~\ref{fig:case2} shows the VP for varying $\beta $ for the same black holes as in Fig.~\ref{fig:L8_15DirNeu}. 
As in previous figures, these plots indicate that, far from the black hole, the VP approaches the Neumann vacuum value in pure adS space-time (\ref{eq:adSN}), except when $\beta =0$ and the VP approaches the Dirichlet vacuum value (\ref{eq:adSD}). 

Finally, in Fig.~\ref{fig:case3} we show the VP for varying $\beta $ for high-temperature black holes.
In this case the VP for the thermodynamically unstable spherical black holes with $k=1^{(-)}$ is many times larger than the VP for the remaining black holes ($k=1^{(+)}$, $k=0$ and $k=-1$). 
Therefore in this figure we do not show the VP for the $k=1^{(-)}$ black holes. 
For the black holes in Fig.~\ref{fig:case3}, both the event horizon radius $r_{h}=0.0101$ and adS radius of curvature $L\approx 0.001$ are comparatively small, but the surface gravity $\kappa \approx 14801$ is very large. 
Due to the high temperature, we find very large values of the VP everywhere. For spherical and hyperbolic black holes, the divergence of the VP on the horizon as $\beta \rightarrow \beta _{\rm {crit}}$ is particularly marked in this case. 
The qualitative features of the VP are similar to those shown in previous plots. 
In particular, it appears that far from the black hole the VP for all $\beta $ other than $\beta =0$ approaches the vacuum value in pure adS space-time (\ref{eq:adSN}) with Neumann boundary conditions applied.

\section{Conclusions}
\label{sec:conc}

We have computed the renormalized VP for a massless, conformally coupled, scalar field on topological, asymptotically-adS, black holes.
The event horizon may have positive, negative or zero curvature, corresponding to spherical, hyperbolic or planar surfaces.
Hyperbolic and planar black holes are always thermodynamically stable.
Spherical black holes exist only for temperatures above a minimum value, above which there are two branches of spherical black holes: larger black holes are also thermodynamically stable but smaller black holes are thermodynamically unstable.

The scalar field satisfies Robin boundary conditions, parameterized by an angle $\alpha $.
For all event horizon topologies, there is a critical value of $\alpha $, above which the field has classically unstable modes. 
This critical value of $\alpha $ depends on the radius and curvature of the event horizon of the black hole. 
We therefore consider the quantum scalar field only for values of $\alpha $ for which there are no classical instabilities.

In order to compute the VP, we employ the ``extended coordinates'' method of \cite{Taylor:2016edd,Taylor:2017sux,Morley:2018lwn}.
This approach enables us to perform the required renormalization mode-by-mode, so that the renormalized VP is written as a mode sum which converges rapidly.
The modes themselves are computed numerically. 
The modes with Robin boundary conditions applied are written as linear combinations of those for Dirichlet and Neumann boundary conditions, which saves computational effort.
We have presented results for the renormalized VP for a variety of topological black holes (including small black holes, large black holes, low and high temperatures), as a function of the parameter governing the boundary conditions and the distance from the horizon.

In previous work \cite{Morley:2018lwn}, we computed the VP when Dirichlet boundary conditions are applied to the scalar field.
We found that for all the black holes considered, the VP was monotonically decreasing from its value on the event horizon to that at infinity.
On the spacetime boundary, the VP always approached the vacuum value on pure adS with Dirichlet boundary conditions applied. 
We also found that, for planar and hyperbolic black holes the VP was negative everywhere on and outside the event horizon.
However, for black holes with spherical event horizons, small black holes have a VP which is positive on the horizon, while the VP for larger spherical black holes was negative everywhere. 
We conjectured that the sign of the VP on the event horizon might be related to the thermodynamic stability of the black hole, with thermodynamically stable black holes having a VP which was negative everywhere.

In this paper we have extended the results of \cite{Morley:2018lwn} by applying mixed (Robin) boundary conditions to the scalar field. 
This work was motivated by our recent study \cite{Morley:2020ayr} of the VP for vacuum and thermal states on pure adS with Robin boundary conditions. 
The first key result of \cite{Morley:2020ayr} concerned the behaviour of the VP on the spacetime boundary.
In particular, for both vacuum and thermal states, and all boundary conditions other than Dirichlet boundary conditions, we found that the VP approached the same asymptotic limit.
The second key result of \cite{Morley:2020ayr} was that as the parameter $\alpha $ in the Robin boundary conditions approached $\alpha _{\rm {crit}}$ (the value of $\alpha $ above which there are classical instabilities), the VP diverged, indicating a breakdown in the semiclassical approximation. 

Those two key results for the VP on pure adS are replicated here for the VP on topological black hole spacetimes.
In all cases studied here, we find that, as the critical value of the parameter $\alpha $ describing the Robin boundary conditions is approached, the black hole VP diverges on the horizon, again indicating that the semiclassical approximation ceases to be valid when there are classical instabilities.  
On the spacetime boundary, with the scalar field satisfying Robin boundary conditions, we also find that the black hole VP approaches the pure adS vacuum value for Neumann boundary conditions, except when Dirichlet boundary conditions are applied. 
Thus the behaviour of the VP on topological black hole spacetimes, as the boundary is approached, is the same as on pure adS spacetime. 
This result applies irrespective of the event horizon topology or the temperature of the black hole.

The qualitative features of the VP as a function of the distance from the event horizon depend on the black hole temperature and the boundary conditions. 
In contrast to the situation for Dirichlet and Neumann boundary conditions, in general the VP is not necessarily a monotonic function of the radial coordinate. 
The rate at which the VP approaches its value on the boundary depends on both the boundary conditions and the temperature.
The VP for higher-temperature black holes converges more slowly as a function of radius than that for lower-temperature black holes. 
The VP also converges more slowly as the parameter $\alpha $ approaches its critical value.
Our final result is that the sign of the VP also depends on the black hole event horizon radius and the temperature. 
Contrary to the conjecture put forward in \cite{Morley:2018lwn} for Dirichlet boundary conditions, for general Robin boundary conditions there appears to be no simple correlation between the sign of the VP on the event horizon and the thermodynamic stability of the black hole.

In this paper we have considered the simplest expectation value for a quantum scalar field, namely the renormalized VP, and have also restricted our attention to a massless, conformally coupled, scalar field.
It would be very interesting to extend this work to massive scalar fields, or other couplings to the curvature, as well as to the object of primary interest in quantum field theory in curved spacetime, the RSET.  
Both these generalizations will likely involve significant technical challenges.
 
\begin{acknowledgments}
T.M.~thanks the School of Mathematics and Statistics at the University of Sheffield for the provision of a studentship supporting this work. 
The work of E.W.~is supported by the Lancaster-Manchester-Sheffield Consortium for Fundamental Physics under STFC grant ST/P000800/1 and partially supported by the H2020-MSCA-RISE-2017 Grant No.~FunFiCO-777740. 
\end{acknowledgments}

\bibliography{MTWrefs}

\end{document}